\documentclass[10pt,journal,compsoc]{IEEEtran}

\ifCLASSOPTIONcompsoc
  \usepackage[nocompress]{cite}
\else
  \usepackage{cite}
\fi

\newcommand{\sofa}{\textsc{gSoFa}}
\usepackage{xcolor}
\usepackage{mathtools}
\usepackage{amsmath}
\usepackage{url}
\usepackage{setspace}
\usepackage{colortbl}
\usepackage[utf8]{inputenc}
\usepackage[english]{babel}
 \usepackage[switch]{lineno}
 \usepackage{tikz} 
 \usepackage{amsfonts}
 \usepackage[font=small]{caption}
\usepackage{floatrow} 
 \usepackage{comment}
\usepackage{color,soul}
\usepackage{textcomp}

\newtheorem{theorem}{Theorem}
\newtheorem{definition}{Definition}

\definecolor{mygreen}{rgb}{0.0, 0.6, 0.0}
\definecolor{myblue}{rgb}{0.0, 0.0, 1.0}
\definecolor{mypink}{rgb}{1.0, 0.0, 1.0}
\newcommand{\ignore}[1]{}

             \usepackage{multirow}

\usepackage{tikz}

\newcommand*\circled[1]{\tikz[baseline=(char.base)]{
            \node[shape=circle,fill,inner sep=1pt] (char) {\footnotesize \textcolor{white}{#1}};}}

 \usepackage{graphicx}   
 \usepackage{subfig}
 \usepackage{wrapfig,lipsum,booktabs}
 \newcommand*{\tikzbullet}[2]{%
   \setbox0=\hbox{\strut}%
   \begin{tikzpicture}
     \useasboundingbox (-.25em,0) rectangle (.25em,\ht0);
     \filldraw[draw=#1,fill=#2] (0,0.5\ht0) circle[radius=.25em];
   \end{tikzpicture}%
}

\begin{document}


\title{{\sofa}: Scalable Sparse Symbolic\\ LU Factorization on GPUs}

\author{Anil~Gaihre,~\IEEEmembership{}
        Xiaoye~Sherry~Li~\IEEEmembership{}
        and~Hang~Liu~\IEEEmembership{}
\IEEEcompsocitemizethanks{\IEEEcompsocthanksitem A. Gaihre and H. Liu are with Department
of Electrical and Computer Engineering, Stevens Institute of Technology, NJ.\protect\\
E-mail: \{agaihre, hliu77\}@stevens.edu
\IEEEcompsocthanksitem X.S. Li is with Lawrence Berkeley National Laboratory, Berkeley, CA.\protect\\
E-mail: xsli@lbl.gov}
\thanks{Manuscript revised May 7, 2021.}}

\markboth{Journal of \LaTeX\ Class Files,~Vol.~14, No.~8, August~2015}%
{Shell \MakeLowercase{\textit{et al.}}: Bare Advanced Demo of IEEEtran.cls for IEEE Computer Society Journals}
%



\IEEEtitleabstractindextext{%
\begin{abstract}
Decomposing a matrix $\mathbf{A}$ into a lower matrix $\mathbf{L}$ and an upper matrix $\mathbf{U}$, which is also known as LU decomposition, is an essential operation in numerical linear algebra.
For a sparse matrix, LU decomposition often introduces {more nonzero entries in the $\mathbf{L}$ and $\mathbf{U}$ factors than in the original matrix}. A \textit{symbolic factorization} step is needed to identify the nonzero structures of $\mathbf{L}$ and $\mathbf{U}$ matrices. Attracted by the enormous potentials of the Graphics Processing Units (GPUs), an array of efforts have surged to deploy various LU factorization steps except for the symbolic factorization, to the best of our knowledge, on GPUs. 
{This paper introduces {\sofa}, the first \underline{G}PU-based \underline{s}ymb\underline{o}lic \underline{fa}ctorization design with} the following three optimizations to enable scalable LU symbolic factorization for \textit{nonsymmetric pattern} sparse matrices on GPUs. First, we introduce a novel fine-grained parallel symbolic factorization algorithm that is well suited for the \textit{Single Instruction Multiple Thread} (SIMT) architecture of GPUs. 
 Second, we tailor supernode detection into a SIMT friendly process and strive to {balance the workload}, {minimize the communication} and {saturate the GPU computing resources} during supernode detection. Third, we introduce a three-pronged optimization to reduce the excessive space consumption problem faced by multi-source concurrent symbolic factorization. Taken together, {\sofa} achieves up to 31$\times$ speedup from 1 to 44 Summit nodes (6 to 264 GPUs) and outperforms the state-of-the-art CPU project, on average, by 5$\times$. Notably, {\sofa} also achieves {up to 47\%} of the peak memory throughput of a V100 GPU in Summit. 
\end{abstract}

}

\maketitle

\IEEEdisplaynontitleabstractindextext

%
\IEEEpeerreviewmaketitle


\section{Introduction}

Many scientific and engineering problems require solving large-scale linear systems, i.e., $\mathbf{Ax = b}$. Solving this problem with direct methods~\cite{davis2016survey} often involves LU factorization~\cite{grigori2007parallel,lezar2010gpu}, that is, decomposing the original matrix $\mathbf{A}$ into lower and upper triangular matrices $\mathbf{L}$ and $\mathbf{U}$, respectively, where $\mathbf{A = LU}$. Since LU decomposition of a sparse matrix typically introduces new nonzeros, also known as \emph{fill-ins}, in the $\mathbf{L}$ and $\mathbf{U}$ factored matrices,
\emph{symbolic factorization}~\cite{grigori2007parallel} is used to compute the locations of
the fill-ins for both $\mathbf{L}$ and $\mathbf{U}$ matrices.
This information is needed to allocate the compressed sparse data structures for
$\mathbf{L}$ and $\mathbf{U}$, and for the subsequent numerical factorization.

\begin{figure}[t]
	\floatbox[{\capbeside\thisfloatsetup{capbesideposition={right,top},capbesidewidth=3cm}}]{figure}[\FBwidth]
	{
		\caption{(a) A sparse matrix $\mathbf{A}$ used throughout this paper where each \protect\tikzbullet{black}{black} represents a nonzero, (b) G($\mathbf{A}$): the graph representation of $\mathbf{A}$.
			}
	\label{fig:matrix_begin_end}
	}
	{
		\includegraphics[width=1.06\linewidth]{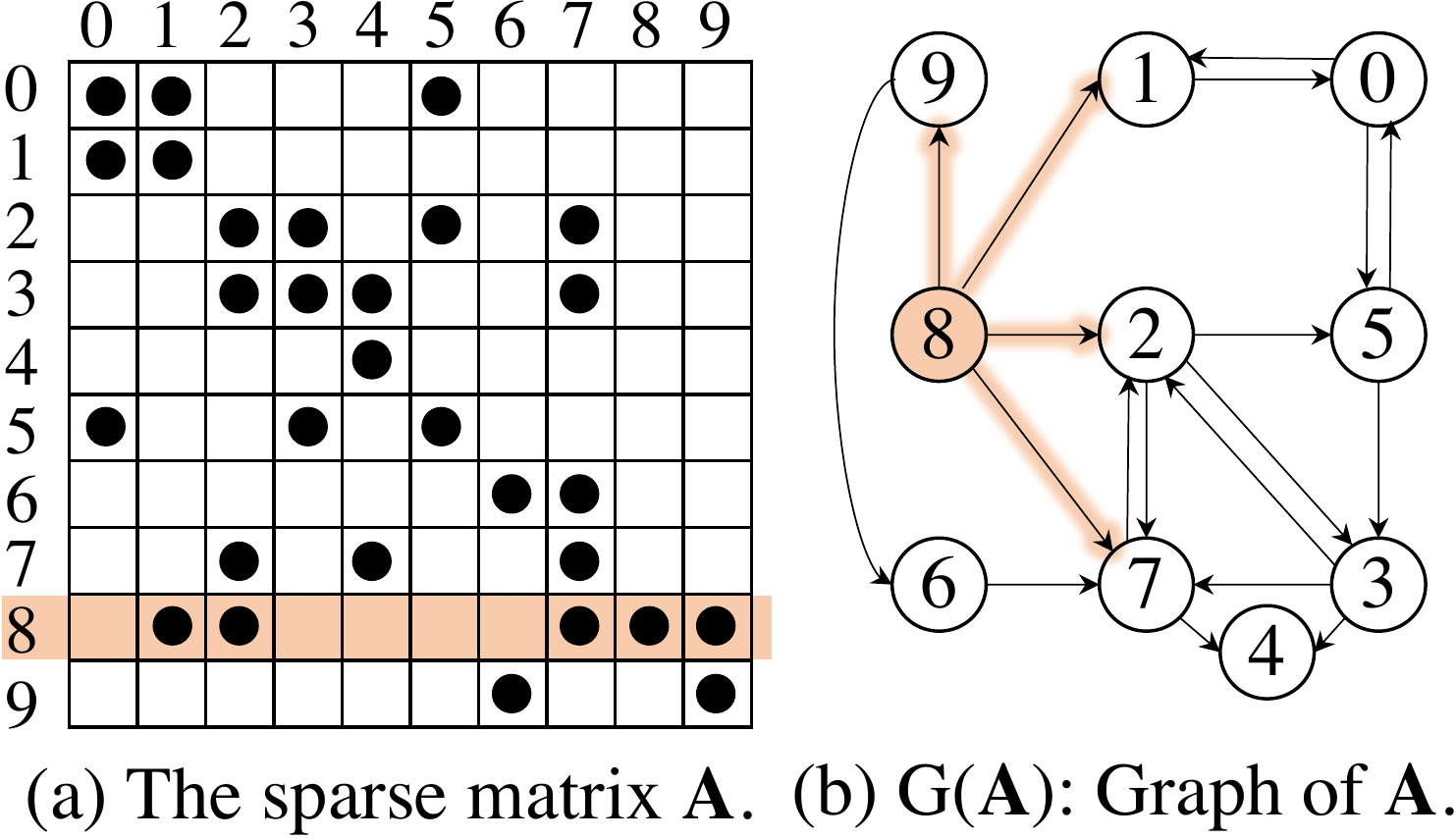}
		\vspace{-.4in}
	}
\end{figure}

Symbolic factorization is a graph algorithm that acts on the adjacency graph
of the corresponding sparse matrix.
Figure~\ref{fig:matrix_begin_end} shows a sparse matrix $\mathbf{A}$
and its adjacency graph $G(\mathbf{A})$.
In the graph, each row of a matrix corresponds to a vertex in the
directed graph with a nonzero in that row corresponding to an out edge of
that vertex.
For instance, row 8 of $\mathbf{A}$ in Figure~\ref{fig:matrix_begin_end}(a)
has nonzeros in columns \{1, 2, 7, 8, 9\}. We thus have those corresponding
out edges for vertex 8 in $G(\mathbf{A})$ of Figure~\ref{fig:matrix_begin_end}(b). 
For brevity, we do not represent the self edges corresponding to the diagonal elements. We will use this example throughout this paper.

The motivation to parallelize the sparse symbolic factorization on GPUs is threefold. Firstly, the newer generation GPUs have a relatively large amount of device memory, which allows the entire sparse matrix to reside on the GPUs. Secondly, more and more exascale application codes are moving to GPUs, which demands the underlying linear solvers to reside on GPUs to eliminate CPU-GPU communication. Ultimately, we will move all the workflows of the sparse LU solver to GPUs.
Third, many GPU-based efforts have gone to deploying numerical factorization and triangular solution on GPUs~\cite{sao2014distributed,liu2016synchronization} due to (i) numerical factorization and triangular solution take more time, and (ii) symbolic factorization consumes excessive memory space. We argue that since sparse symbolic factorization detects the fill-ins and supernodes for the numerical steps, it is imperative to move sparse symbolic factorization on GPUs.

Existing symbolic factorization algorithms are difficult to deploy on GPUs,
as well as scale-up in distributed settings, due to stringent data dependency, lack of
fine-grained parallelism, and excessive memory consumption. There mainly
exist three types of algorithms to perform
symbolic factorization, that is, fill1 and fill2 algorithms by Rose and
Tarjan~\cite{rose1978algorithmic} and
Gilbert-Peierls algorithm~\cite{gilbert1993elimination,gilbert1988sparse}.
They are based on traversing either the graphs
$G(\mathbf{L})$ and $G(\mathbf{U})$ or graph $G(\mathbf{A})$. In particular, fill1 works on $G(\mathbf{U})$ for fill-ins detection while Gilbert-Peierls algorithm does that on $G(\mathbf{L})$.
Fill2 algorithm is analogous to Dijkstra's algorithm~\cite{dijkstra1959note}.
Consequently, fill1 and Gilbert-Peierls algorithms are sequential in nature as the fill-in detection of larger rows or columns has to wait until the completion of the smaller rows or columns. 
Fill2 algorithm lacks fine-grained parallelism (i.e., in each source) due to strict priority requirement, although coarse-grained parallelism (i.e., at source level) exists. Furthermore, symbolic factorization needs to identify supernode~\cite{demmel1999supernodal}, which is used to expedite the numerical steps. But supernode detection presents cross-source data dependencies. Third, we need to perform parallel symbolic factorization for a collection of rows to saturate the computing resources on GPUs. This results in overwhelming space consumption, which is similar to multi-source graph traversal~\cite{liu2016ibfs}.

In this paper, we choose the fill2 algorithm as a starting point stemming from the fact that fill2 directly traverses the original instead of the filled graphs, allowing independent
traversals from multiple sources. However, we have to revamp this algorithm
to expose a higher degree of fine-grained parallelism to match GPUs' SIMT nature and reduce the excessive memory requirements during the multi-source symbolic factorization to, again, fit GPUs, which often equip limited memory space.
To summarize, we make the following three major contributions.

First, to the best of our knowledge, we introduce the first fine-grained
parallel symbolic factorization that is well suited for the SIMT architecture of GPUs. Particularly, instead of processing one vertex at a time, we allow all the neighboring vertices
to be processed in parallel, such that the computations align well with the SIMT nature.
\textit{Since this relaxation enables traversal to some vertices before their dependencies are satisfied,  we further allow revisitation of these vertices to ensure correctness}. We also introduce optimizations to
reduce the reinitialization overhead and random memory access to the algorithmic data.
{Finally, we use multi-source concurrent symbolic factorization to saturate each GPU and achieve intra-GPU workload balance.}

Second, we not only tailor supernode detection into a SIMT friendly process but also strive to \textit{balance the workload}, \textit{minimize the communication} and \textit{saturate GPU computing resources} during supernode detection. Particularly, we break supernode detection into two phases to expose massive parallelism for GPUs. Further, we assign a chunk of continuous sources to one computing node to avoid inter-node communication and interleave the sources of each chunk across GPUs in each node to balance the workload. Eventually, we investigate the configuration space of unified memory and propose a design that can significantly reduce the page faults during supernode detection.

Third, we introduce a three-pronged optimization to combat the excessive space consumption problem faced by multi-source concurrent symbolic factorization:
(i) We propose the external GPU frontier management because the space requirement of frontier-related data structures is very dynamic, and their access pattern is predictable.
(ii) We identify and remove the ``bubbles'' in vertex status-related data structures;
and (iii) since various data structures present dynamic space requirements with respect
to different sources, we propose allocating single memory space for all these data structures and dynamically adjust their capacities to reduce the frequency of copying frontier-related data structures between CPU and GPU memories.


The rest of this paper is organized as follows: Section~\ref{sec:background} introduces the background of this work. Section~\ref{sec:challenge} presents the design challenges.  {Sections~\ref{sec:parallel},~\ref{sec:supernode} and~\ref{sec:space} present the fine-grained symbolic factorization algorithm design, parallel efficient supernode detection, and space consumption optimization techniques.} We evaluate {\sofa} in Section~\ref{sec:eval}, study the related work in Section~\ref{sec:related}, and conclude in Section~\ref{sec:conclusion}.

\section{Background}\label{sec:background}



\subsection{Sparse LU Factorization}
\label{subsec:back:lu}

Factorizing a sparse matrix $\mathbf{A}$ into the $\mathbf{L}$ and $\mathbf{U}$ matrices often involves several major steps, such as preprocessing, symbolic factorization, and numerical factorization. 

\textbf{Matrix preprocessing} performs either row (partial pivoting) or both row and column permutations (complete pivoting) in order to
\textit{improve numerical stability} and
\textit{reduce the number of fill-ins} in the $\mathbf{L}$ and $\mathbf{U}$ matrices. For numerical stability, existing methods aim to find a
permutation so that the pivots on the diagonal are big, for instance,
by maximizing the product of diagonal entries and make the matrix diagonal
dominant~\cite{duff1999design}. {Supplemental file illustrates how matrix preprocessing works with matrix $\mathbf{A}$.}
Regarding fill-in reduction, the popular strategies are {minimum degree algorithm} \cite{tinney1967direct}, {Reverse Cuthill-McKee }\cite{cuthill1969reducing} and {nested dissection}~\cite{kernighan1970efficient}, or a combination of them, such as METIS~\cite{karypis1998parallel}. 

\begin{figure}[t]
	\centering
	\includegraphics[width=.98\textwidth]{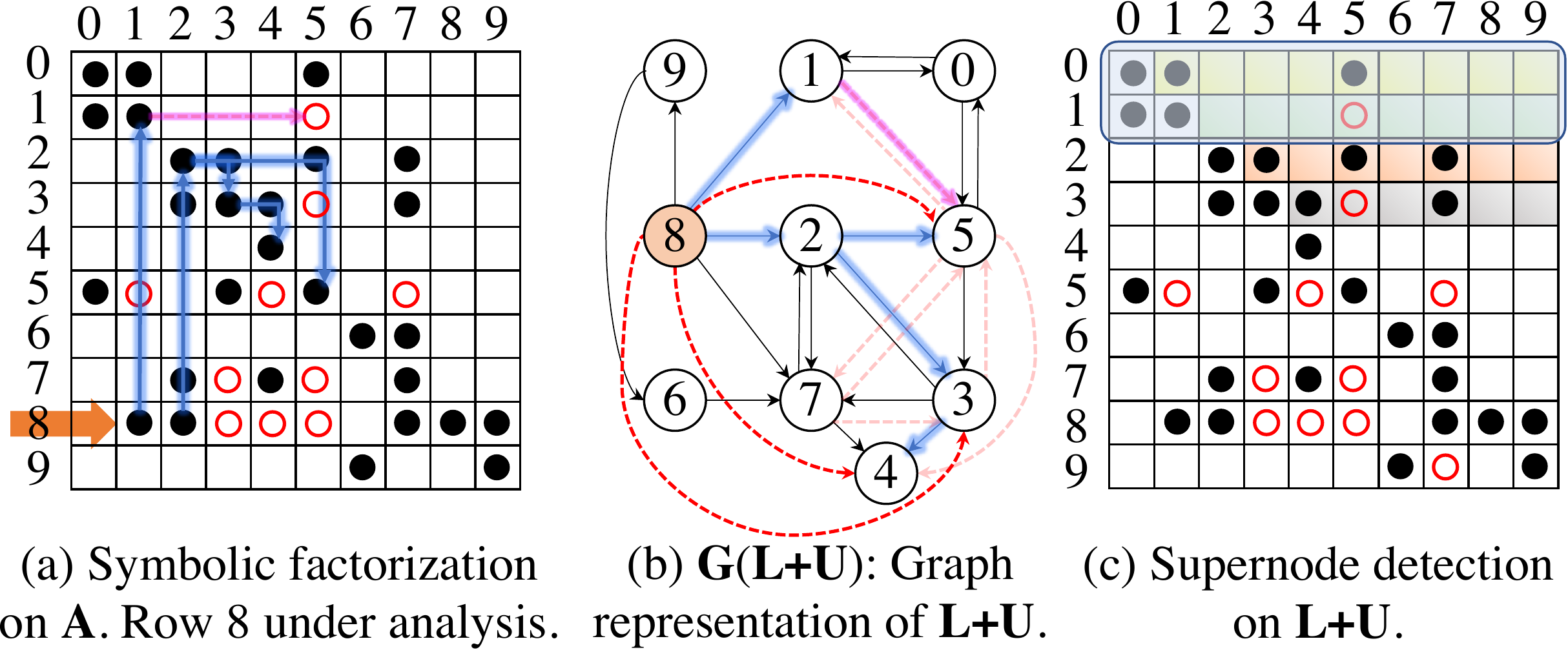} 
\vspace{-.05in}
		\caption{(a) Symbolic factorization on sparse matrix with row 8 under analysis \textbf{A}, (b) the Graph representation of filled matrix \textbf{(L+U)} with row 8 under analysis and (c) Supernode detection on filled matrix.  The \protect\tikzbullet{red}{white} is the new nonzeros (fill-ins) in the \textbf{L} and \textbf{U} factors. And \textcolor{red}{$\dashrightarrow$}/\textcolor{mypink}{$\dashrightarrow$}  indicates the new edges in the $G(\mathbf{L}+\mathbf{U})$ graph. 
	{The dark edges with blue or pink glows in (a) and (b) show the traversal paths that lead to new fill-ins.} 
	  \vspace{-.25in}
	  }
	\label{fig:matrix_symbolic} 
\end{figure}

\textbf{Symbolic factorization.} 
Symbolic factorization can be viewed as performing Gaussian elimination on a
sparse matrix with either 0 or 1 entries. At each elimination step, the pivoting row is multiplied by a nonzero scalar and updated into another row below, which has nonzero in the pivoting column; this is called an elementary row operation.
After a sequence of row operations, the original
matrix is transformed into two triangular matrices $\mathbf{L}$ and $\mathbf{U}$.
Considering row 8 in Figure~\ref{fig:matrix_symbolic}(a), since row 8 has nonzeros at columns 1 and 2, both rows 1 and 2 will perform row operations on row 8.
Using row operation from row 2 as an example, it produces a fill-in (8, 3), which further triggers row operations from rows 3 and 5 that produce fill-ins (8, 4) and (8, 5). We will detail this process shortly. 
The following fill-path theorem succinctly characterizes the fill-in locations
reflecting the above elimination process.

\begin{theorem}
\label{thm:fill}
A fill-in at index (i, j) is introduced if and only if there exists a directed path from i to j, with the intermediate vertices being smaller than both i and j~\cite{rose1978algorithmic}.
\end{theorem}

Theorem~\ref{thm:fill} can be applied on either the (partially) filled graph $G(\mathbf{U})$ by the fill1 algorithm or the original graph $G(\mathbf{A})$ by the fill2 algorithm~\cite{rose1978algorithmic}.

\begin{figure}[t]
	\centering
	\includegraphics[width=1\textwidth]{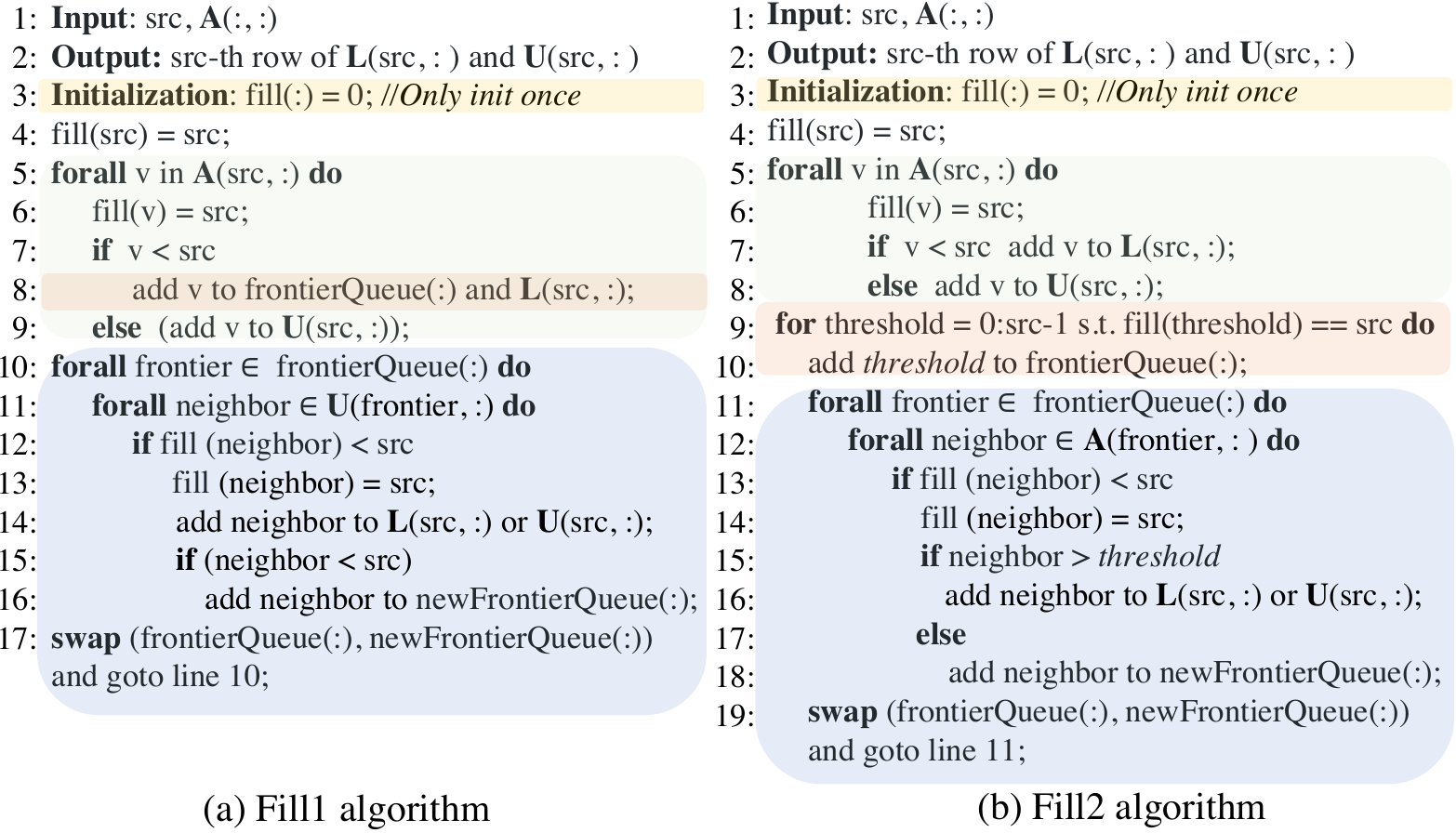} 
	\caption{Fill1 and fill2 algorithms. We use MATLAB format to denote both arrays and matrices. The snippet with the same background colors in the algorithms represent equivalent sections of these two algorithms.
	\vspace{-.2in}
	} 
	\label{fig:alg_back} 
\end{figure}

Figure~\ref{fig:alg_back}(a) shows how fill1 works for the \textit{src}-th row in three steps: (i) lines 3-9 initialize the fill(:) and frontierQueue(:). While the usages of frontierQueue(:) and newFrontierQueue(:) are straightforward, fill(:) is a flag array that tracks which vertex is visited by src vertex. (ii) If the neighbor vertex is first-time visited (i.e., lines 13-16), we mark this neighbor as visited by setting fill(neighbor) = src at line 13, and add \textit{neighbor} to either $\mathbf{L}$(src, :) or $\mathbf{U}$(src, :) based upon the location of (src, neighbor). (iii) If \textit{neighbor} is smaller than \textit{src}, we add it to \textit{newFrontierQueue(:)} for next iteration traversal.
Figure~\ref{fig:matrix_symbolic}(b) shows how row 8 works under fill1 algorithm. That is, 8 \textcolor{myblue}{$\rightarrow$} 2\textcolor{myblue}{$\rightarrow$} 3 leads to (8, 3),
8 \textcolor{myblue}{$\rightarrow$} 2 \textcolor{myblue}{$\rightarrow$} 3 \textcolor{myblue}{$\rightarrow$} 4 leads to (8, 4),
8 \textcolor{myblue}{$\rightarrow$} 1 \textcolor{mypink}{$\dashrightarrow$} 5 leads to (8, 5)
. Note that the
third path goes through a new fill-in edge (1,5).

Figure~\ref{fig:alg_back}(b) presents the fill2 algorithm. Similar to fill1, it uses fill(:) array to indicate an already visited vertex by setting fill(neighbor) = src. 
The algorithm differs from fill1 mainly on two points. (i) It uses the original graph G($\mathbf{A}$) for traversal (line 12). Second, during traversal, this algorithm only permits traversing one vertex (i.e., threshold) at a time, starting from the smallest one (line 9 - 10). For each threshold that is treated as a frontier, this algorithm checks its neighbors, updates the statuses of the neighbors in fill(:), and adds new fills to either $\mathbf{L}$(src,:) or $\mathbf{U}$(src,:), as well as to the newFrontierQueue(:) if this neighbor obeys Theorem~\ref{thm:fill}. This process continues until the vertices that are smaller and connected to the threshold vertex are exhausted, i.e., frontierQueue(:) is empty. Subsequently, fill2 will proceed to the next threshold vertex in line 9.  
Considering row 8 in Figure~\ref{fig:matrix_symbolic}(b) again, the three fill-ins are due to the three paths going through only existing edges: 8 \textcolor{myblue}{$\rightarrow$} 2\textcolor{myblue}{$\rightarrow$} 3 leads to (8, 3),
8 \textcolor{myblue}{$\rightarrow$} 2 \textcolor{myblue}{$\rightarrow$} 3 \textcolor{myblue}{$\rightarrow$} 4 leads to (8, 4),
8 \textcolor{myblue}{$\rightarrow$} 2 \textcolor{myblue}{$\rightarrow$} 5 leads to (8, 5).

A decade later,~\cite{gilbert1988sparse} introduces the Gilbert-Peierls algorithm to find the fill-in structures, which is a simpler way of interpreting fill1 algorithm. 
Specifically, this approach determines the nonzero structures column by column.
For clarity, we define that $\mathbf{L}(i:j,m:n)$ and $\mathbf{U}(i:j,m:n)$
denote the block from rows $i$ to $j$, and columns $m$ to $n$ in $\mathbf{L}$
and $\mathbf{U}$ matrices, respectively. For column $k$, it traverses the
graph $\mathbf{L}(:,0:k-1)^{T}$ in a Depth-First Search (DFS) manner. The vertex that is reachable by the vertices in column $k$ results
in a fill-in at column $k$. {The graph used by~\cite{gilbert1988sparse} is
  similar to that of fill1 except that~\cite{gilbert1988sparse} applies a
  transpose to the $\mathbf{L}$ matrix. Further, the vertices that
  can be reached from the source vertices will
  automatically satisfy Theorem~\ref{thm:fill} because only the graph of
  $\mathbf{L}(:,0:k-1)^{T}$ is used.}

It is worth noting that the nonsymmetric-pattern sparse LU symbolic factorization is much harder than the symmetric-pattern counterpart: (i) In the symmetric case, the transitive reduction of the filled graph $G(\mathbf{L})$ is a tree, called elimination tree; symbolic factorization using the elimination tree can be done in time
$O(nnzeros(\mathbf{L}))$, linear to the output size.
(ii) However, for the nonsymmetric cases, the transitive reduction of the filled graphs $G(\mathbf{L})$ and $G(\mathbf{U})$ are Directed Acyclic Graphs (DAGs), called elimination DAGs. Computing these DAGs is expensive, and all variants of this method on the nonsymmetric symbolic factorization algorithms take asymptotically longer than linear time~\cite{gilbert1988sparse,rose1978algorithmic}. 

\textbf{Numerical factorization.}
After the structure of the fill-ins is determined, the solvers perform numerical factorization to calculate the values of the $\mathbf{L}$ and $\mathbf{U}$ matrices. 
Popular numerical factorization methods are \textit{left-looking}~\cite{gilbert1988sparse} and \textit{right-looking}~\cite{grigori2007parallel}. 
{We explain how numerical factorization works on matrix $\mathbf{A}$ in supplemental file}.

\textbf{Triangular solution.} Once LU factorization arrives at $\mathbf{A}=\mathbf{LU}$, we can solve $\mathbf{Ax}=\mathbf{b}$ in two steps with triangular solver, given $\mathbf{Ax}=\mathbf{b}$ becomes $\mathbf{LUx}=\mathbf{b}$. First, triangular solver can solve $\mathbf{Ly} = \mathbf{b}$ to derive $\mathbf{y}$. Second, triangular solver derives $\mathbf{x}$ by solving $\mathbf{Ux} = \mathbf{y}$.

\subsection{{Supernode}}
Symbolic factorization also needs to identify the supernodes in $G(\mathbf{L}+\mathbf{U})$, which is important to improve the performance for both numerical factorization and triangular solver.
Particularly, \textit{a supernode is a range of rows or columns with the same nonzero structures}, such as rows 1 and 2 in Figure~\ref{fig:matrix_symbolic}(c), so that these rows or columns can be treated as a  dense matrix~\cite{demmel1999supernodal}.
During numerical factorization, a dense matrix format allows us to use Level 3 Basic Linear Algebra Subprograms (BLAS) operations such as matrix-matrix multiplication, 
which is often faster than the lower level BLAS operations like Level 2 BLAS operations. In this paper, {\sofa} supports T3, one of the most popular types of supernodes among five types of supernodes~\cite{demmel1999supernodal}. Definition~\ref{def:T3} defines a T3 supernode. 

\begin{definition}
\label{def:T3}
Supposing a T3 supernode begins at row $r$ and extends through row $s-1$, row $s$ belongs to T3 if and only if $nnz(\mathbf{U}(s,:))=nnz(\mathbf{U}(s-1,:))-1$ and $\mathbf{L}(s, r) \neq 0$ according to~\cite{demmel1999supernodal}. 
\end{definition}

Definition~\ref{def:T3} states that for a supernode that already contains rows $r$ through $s-1$, two requirements are needed for the next row $s$ to be included in this supernode: (i) The number of nonzeros in row $s$ of $\mathbf{U}$ shall be one fewer than that of row $s-1$. (ii) There shall be a nonzero at $(s, r)$ in the filled matrix $\mathbf{L}$. If either of the requirements is not met, row $s$ will not belong to the supernode. Intuitively, (i) a nonzero at $\mathbf{L}(s, r)$ ensures that all the nonzero patterns of row $r$ in $\mathbf{U}$ are mapped into row $s$ during symbolic factorization. (ii) Further, because the nonzero count of row $s$ is one fewer than that of row $s-1$ in $\mathbf{U}$, we can conclude rows $s-1$ and $s$ follow the same nonzero pattern in $\mathbf{U}$. Thus, they belong to the same supernode. 


Now, we use the example in Figure~\ref{fig:matrix_symbolic}(c) to illustrate how to detect a supernode in $G(\mathbf{L}+\mathbf{U})$.
Starting from row 0, we check whether (i) the number of nonzeros in row 1
of $\mathbf{U}$ matrix (i.e. 2) is one fewer than row 0 (i.e., 3); and (ii) there is a nonzero at (1, 0). Since both requirements are met, row 1 is included in the supernode starting at row 0. Applying a similar process to row 2, we find that row 2 has neither one less nonzero than the previous row (i.e., row 1) nor a nonzero at (2, 0). Hence, row 2 starts a new supernode.

\subsection{Graphics Processing Units}

This section discusses general-purpose GPUs with
recent NVIDIA V100 GPU~\cite{nvidia2017nvidia} as an example. 

\textbf{Streaming processors and threads.} The V100 GPU is designed with NVIDIA Volta architecture. V100 is powered by 80 Streaming Multiprocessors (SMX). Each SMX features 64 CUDA cores, resulting in a total of 5,120 CUDA cores. During execution, a GPU thread runs on one CUDA core. A SMX schedules a group of 32 consecutive threads known as a warp in a SIMT manner. A collection of consecutive warps further formulate a Cooperative Thread Array (CTA), or a block. All the CTAs together in one kernel are called a grid.
%

\textbf{Memory architecture.} V100 comes with two memory capacities, that is, 16 GB and 32 GB with peak bandwidth up to 900 GB/s. 
Each SMX has 96 KB on-chip fast memory that is shared by the configurable shared memory and L1 cache. All the SMXs share a L2 cache at a size of 6,144 KB. Each thread block can use up to 65,536 32-bit registers. 

\textbf{Fine-grained parallelism}. GPUs favor fine-grained parallelism. Particularly, GPUs can only achieve the aforementioned ideal computing and memory throughput when a warp of threads is working on the same instruction and fetching data from consecutive memory addresses. Otherwise, GPUs might suffer from either warp divergence or uncoalesced memory access issues, resulting in order of magnitude performance degradation~\cite{nvidia2017nvidia}.

\textbf{Unified memory}~\cite{li2019compiler} can unite the memory space of all GPUs and CPUs into a single virtual address space. During execution, any process or thread can access the data from this virtual address space. In the background, the required data is either transferred from where the data resides to where the data is needed implicitly or directly accessed remotely. This is different from the traditional explicit method, which requires programmers to explicitly call \textit{cudaMemcpy()} (or similar functions) in order to transfer the data. Consequently, unified memory bests explicit transfer when the requested data is either small (in terms of size) or randomly accessed, or both. Further, unified memory can be used without terminating the kernel, which is not possible from the explicit method.

\begin{table}[hbt!]
{
\fontsize{8}{10}\selectfont
\begin{center}
\setlength\tabcolsep{1.5pt}
\begin{tabular}{llrrrrrr}


\hline
\multicolumn{1}{|l|}{\textbf{ Matrix ({A})}} & \multicolumn{1}{l|}{\textbf{Abbr.}} & \multicolumn{1}{l|}{\textbf{Order (A)}} & \multicolumn{1}{l|}{\textbf{nnz (A)}} & \multicolumn{1}{l|}{\textbf{\begin{tabular}[c]{@{}l@{}}Struct. \\ symm.\end{tabular}}} &
\multicolumn{1}{l|}{$\frac{\textrm{\bf\fontsize{8}{10}\selectfont nnz(A)}}{\textrm{\bf\fontsize{8}{10}\selectfont Order(A)}}$} &
\multicolumn{1}{l|}{$\frac{\textrm{\bf\fontsize{8}{10}\selectfont \#Fill-in}}{\textrm{\bf\fontsize{8}{10}\selectfont nnz(A)}}$}\\
\hline
\multicolumn{1}{|l|}{BBMAT}& \multicolumn{1}{l|}{BB}    & \multicolumn{1}{r|}{38,744}& \multicolumn{1}{r|}{1,771,722}        & \multicolumn{1}{r|}{0.53}       & \multicolumn{1}{r|}{45.7}   & \multicolumn{1}{r|}{18.29}  \\ \hline
\multicolumn{1}{|l|}{BCSSTK18}        & \multicolumn{1}{l|}{BC}    & \multicolumn{1}{r|}{11,948}& \multicolumn{1}{r|}{149,090} & \multicolumn{1}{r|}{1} & \multicolumn{1}{r|}{12.47}  &  \multicolumn{1}{r|}{6.52} \\ \hline
\multicolumn{1}{|l|}{EPB2} & \multicolumn{1}{l|}{EP}    & \multicolumn{1}{r|}{25,228}& \multicolumn{1}{r|}{175,027} & \multicolumn{1}{r|}{0.67}       & \multicolumn{1}{r|}{6.93} &  \multicolumn{1}{r|}{9.28} \\ \hline
\multicolumn{1}{|l|}{G7JAC200SC}      & \multicolumn{1}{l|}{G7}    & \multicolumn{1}{r|}{59,310}& \multicolumn{1}{r|}{717,620} & \multicolumn{1}{r|}{0.03}       & \multicolumn{1}{r|}{12.1}  & \multicolumn{1}{r|}{24.51}  \\ \hline
\multicolumn{1}{|l|}{LHR71C} & \multicolumn{1}{l|}{LH}    & \multicolumn{1}{r|}{70,304}& \multicolumn{1}{r|}{1,528,092}        & \multicolumn{1}{r|}{0} & \multicolumn{1}{r|}{21.7}  &  \multicolumn{1}{r|}{3.10}   \\ \hline
\multicolumn{1}{|l|}{MARK3JAC}   &
\multicolumn{1}{l|}{MK}    & \multicolumn{1}{r|}{64,089}& \multicolumn{1}{r|}{376,395} & \multicolumn{1}{r|}{0.07}       & \multicolumn{1}{r|}{5.9}   &  \multicolumn{1}{r|}{28.59}  \\ \hline
\multicolumn{1}{|l|}{RMA10}& \multicolumn{1}{l|}{RM}    & \multicolumn{1}{r|}{46,835}& \multicolumn{1}{r|}{2,329,092}        & \multicolumn{1}{r|}{1} & \multicolumn{1}{r|}{49.729}   &  \multicolumn{1}{r|}{3.14} \\ \hline\hline
\multicolumn{1}{|l|}{{AUDIKW\_1}}     & \multicolumn{1}{l|}{{AU}}    & \multicolumn{1}{r|}{{943,695}}      & \multicolumn{1}{r|}{{77,651,847}}        & \multicolumn{1}{r|}{{1}} & \multicolumn{1}{r|}{{82.28}} & \multicolumn{1}{r|}{{31.43}}   \\ \hline
\multicolumn{1}{|l|}{{DIELFILTER}}     &
\multicolumn{1}{l|}{{DI}}    & \multicolumn{1}{r|}{{1,157,456}}      & \multicolumn{1}{r|}{{48,538,952}}        & \multicolumn{1}{r|}{{1}} & \multicolumn{1}{r|}{{41.93}} & \multicolumn{1}{r|}{{22.39}}  \\ \hline
\multicolumn{1}{|l|}{HAMRLE3}& \multicolumn{1}{l|}{HM}    & \multicolumn{1}{r|}{1,447,360}      & \multicolumn{1}{r|}{5,514,242}        & \multicolumn{1}{r|}{0} & \multicolumn{1}{r|}{3.8}    & \multicolumn{1}{r|}{32.63}    \\ \hline
\multicolumn{1}{|l|}{PRE2} & \multicolumn{1}{l|}{PR}    & \multicolumn{1}{r|}{659,033}        & \multicolumn{1}{r|}{5,834,044}        & \multicolumn{1}{r|}{0.33}       & \multicolumn{1}{r|}{8.8}    & \multicolumn{1}{r|}{20.70}   \\ \hline
\multicolumn{1}{|l|}{STOMACH}& \multicolumn{1}{l|}{ST}    & \multicolumn{1}{r|}{213,360}        & \multicolumn{1}{r|}{3,021,648}        & \multicolumn{1}{r|}{0.85}       & \multicolumn{1}{r|}{14.2}   & \multicolumn{1}{r|}{25.77}  \\ \hline
\multicolumn{1}{|l|}{TWOTONE}& \multicolumn{1}{l|}{TT}    & \multicolumn{1}{r|}{120,750}        & \multicolumn{1}{r|}{1,206,265}        & \multicolumn{1}{r|}{0.24}       & \multicolumn{1}{r|}{10}   & \multicolumn{1}{r|}{6.07}    \\ \hline
\end{tabular}
\end{center}
}
\caption{{Dataset specifications.
Note, in graph terminology, order ($\mathbf{A}$) and nnz($\mathbf{A}$) represent $|V|$ and $|E|$ of the graph $G(\mathbf{A})$, respectively, where $\mathbf{A}$ is the matrix of interest.}
} 
\label{tab:datasets}
\end{table}

\subsection{Dataset}
Table~\ref{tab:datasets} presents the datasets that are used to evaluate {\sofa}. They are available from Suite Sparse Matrix Collection~\cite{sparse_matix}.
This dataset collection includes a variety of applications, such as circuit simulation (HM, PR, and TT), structural problems (BC, AU), computational fluid dynamics (BB and RM), thermal problems (EP), economic modeling (G7 and MK), chemical engineering (LH), electromagnetics (DI) and bioengineering problems (ST). 
Besides, this dataset collection covers a wide range of variations in both structural symmetry and sparsity (the fifth and sixth columns in Table~\ref{tab:datasets}). This table arranges the datasets into smaller (upper) and larger (lower) collections with respect to order ($\mathbf{A}$).
The larger matrices are used in space complexity analysis in Sections~\ref{sec:space} and \ref{sec:eval}. Following SuperLU\_DIST~\cite{li2003superlu_dist}, we use ParMETIS~\cite{parmetis} library to preprocess the matrix, and adopt Compressed Sparse Row (CSR) format to represent the matrices~\cite{liu2015enterprise,merrill2012scalable}.

\section{Design Challenges}
\label{sec:challenge}


\subsection*{Challenge \#1. Existing symbolic factorization algorithms present limited fine-grained parallelism.}

Comparing the fill1 and fill2 algorithms, we can see that fill1 is severely limited in parallelism due to data dependency, because, to find the nonzero structure of the current row, we need to wait for the completion of detecting all the nonzero structures of the previous rows in $\mathbf{U}$. In comparison, the fill2 algorithm exhibits a high degree of parallelism:  because graph $G(\mathbf{A})$ is static, we can perform parallel independent traversals from all vertices in $G(\mathbf{A})$. Therefore, an algorithm variant based on fill2 is more favorable for a massively parallel device like GPU.

Despite that fill2 presents more parallelism, fill2 also faces the
limited \textit{fine-grained parallelism issue} because line 9 - 10 of fill2 in Figure~\ref{fig:alg_back}(b) only allows processing one threshold vertex at a time. It is important to note that \textit{simply allowing fill2 to process multiple threshold vertices in parallel will \textit{not} warrant the correctness} because the threshold value has to be strictly incremented sequentially. This is due to the constraint that each vertex can only be visited once in fill2. An example of this argument is discussed in Section~\ref{sec:parallel}.



To mitigate this bottleneck,
{\sofa} allows all frontiers to be processed in parallel, such that a warp of GPU threads can work on consecutive tasks.
\textit{Given this relaxation, some frontiers might be worked on earlier than they are supposed to, we introduce new data structures and control logics to allow revisitations in order to ensure correctness}. 


\subsection*{Challenge \#2. Parallel supernode detection needs a good
  trade-off between load balance and communication reduction.}

\begin{figure}[hbt!]
	\centering
	\vspace{-.25in}
	\begin{tabular}{cc}
		\subfloat[{ BC: small dataset}]{\includegraphics[width=0.4\textwidth]{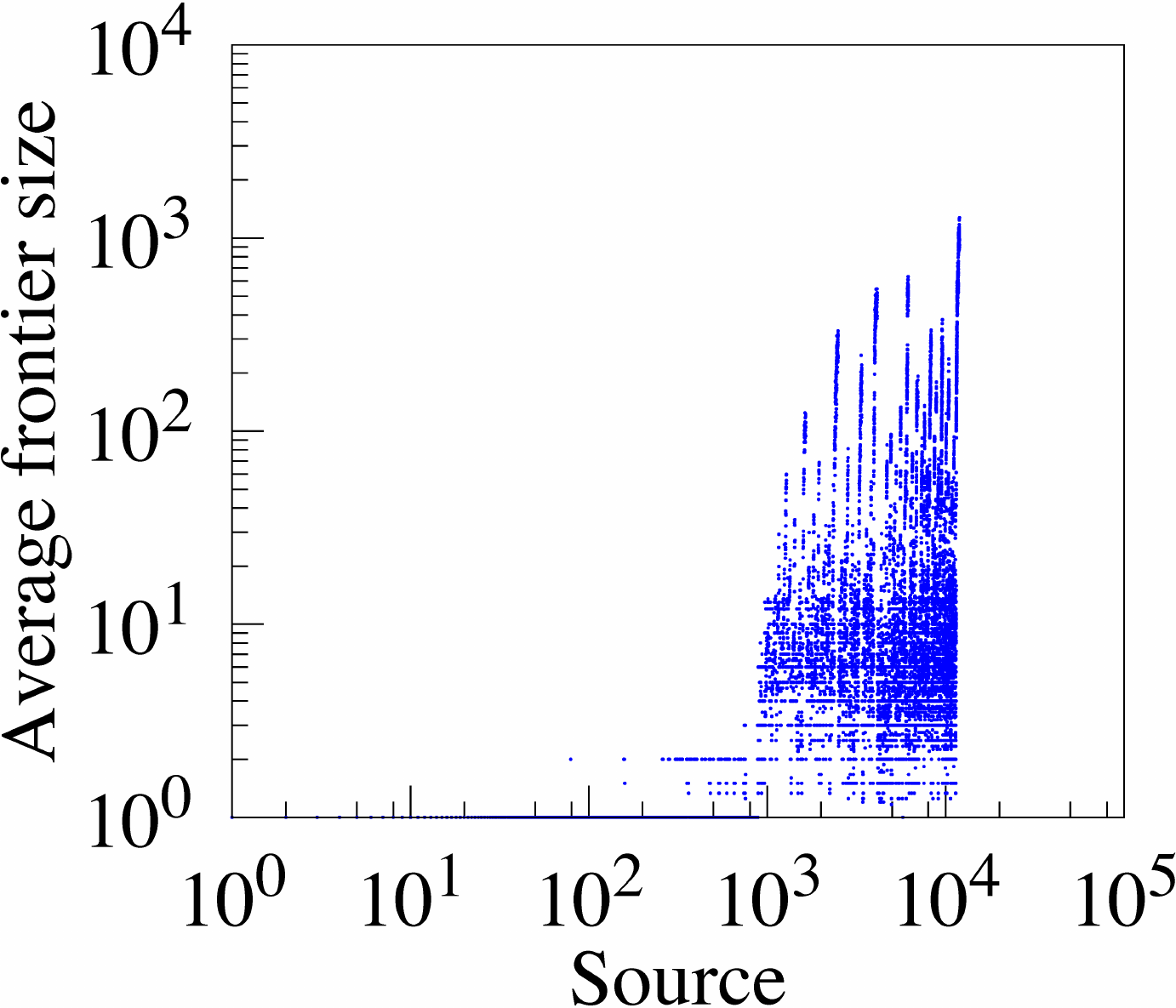}} & 
		\subfloat[{TT: big dataset}]{\includegraphics[width=0.4\textwidth]{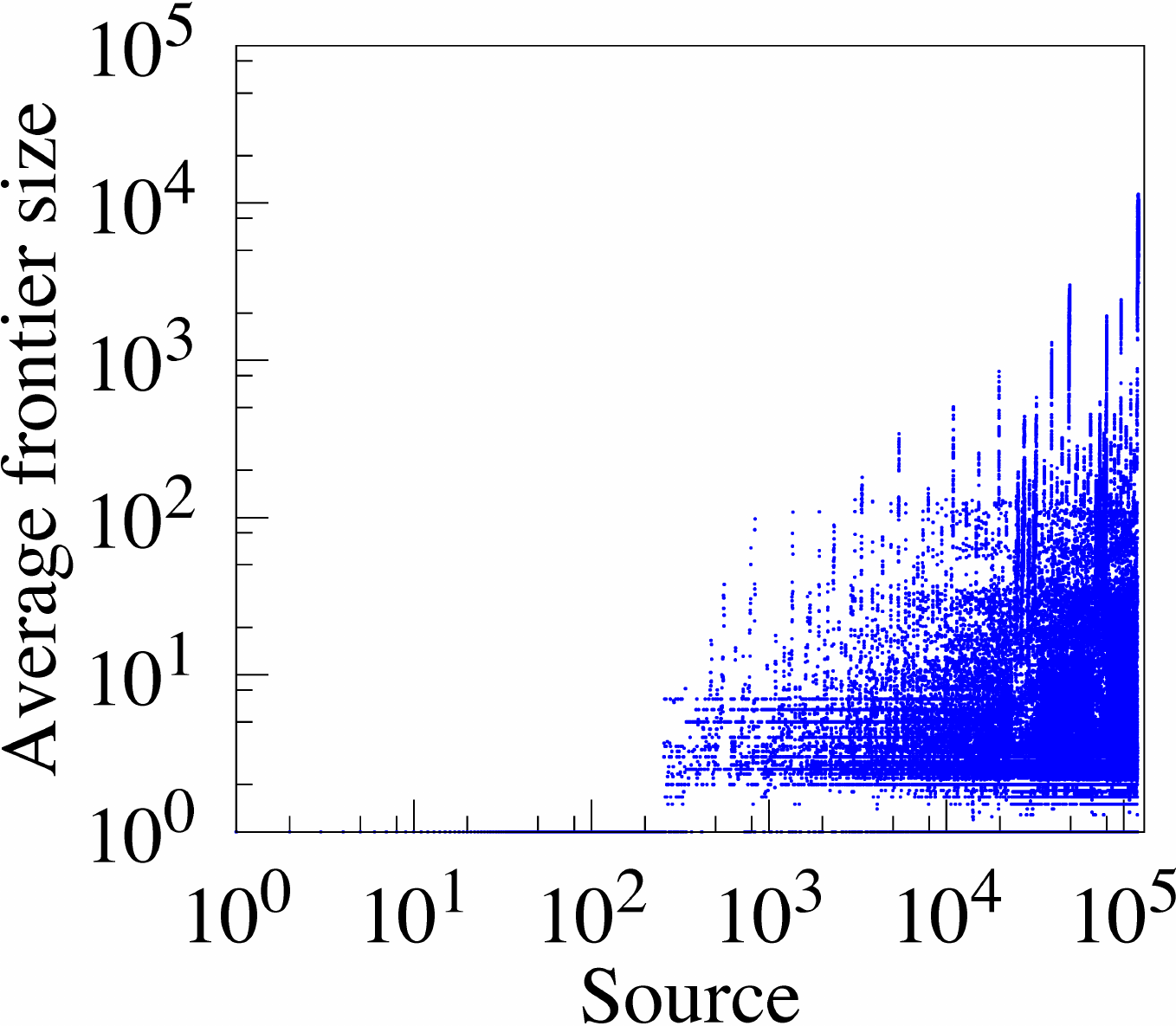}} \\ 
	\end{tabular}
	\vspace{-.12in}
	\caption{Average frontier size per source for {BC and TT}. Note, we only present two datasets for brevity.  
	\vspace{-.1in}
}
	\label{fig:less_workload}
\end{figure}

{Theorem~\ref{thm:fill} indicates that the amount of workload per source generally increases with the increase of source vertex ID,
  because more intermediate vertices will be smaller than the source ID when the source is larger. Our experimental results in
  Figure~\ref{fig:less_workload} corroborate with this.} 
{The general trend is that the workload soars with the increasing source ID.
  For instance, the workload ratios between the smallest and largest sources
  are 1,265$\times$ and 6,230$\times$ for BC and TT datasets, respectively.}

\textit{We note that the optimization to balance the workload
  contradicts the optimization for supernode detection.}
On the one hand, as suggested by the workload dynamic of different sources in Figure~\ref{fig:less_workload}, a proper workload  balancing strategy would require to interleave the sources across GPUs. On the other hand, supernode detection has to check a continuous range of sources, implying that assigning a continuous range of sources to one GPU would minimize the communication cost. These contradictory goals require novel source scheduling and system optimizations.

In this paper, we first transform supernode detection into a massively parallel process that fits GPUs. Subsequently, we introduce a trade-off to both balance the workload and significantly reduce the communication cost during supernode detection with the help of a judicious source scheduling mechanism and an interesting unified memory-based data sharing design.



\subsection*{Challenge \#3. The auxiliary data structures may consume overwhelming memory space.}

\begin{figure}[h]
	\centering
	\vspace{-.05in}
	\includegraphics[width=.96\textwidth]{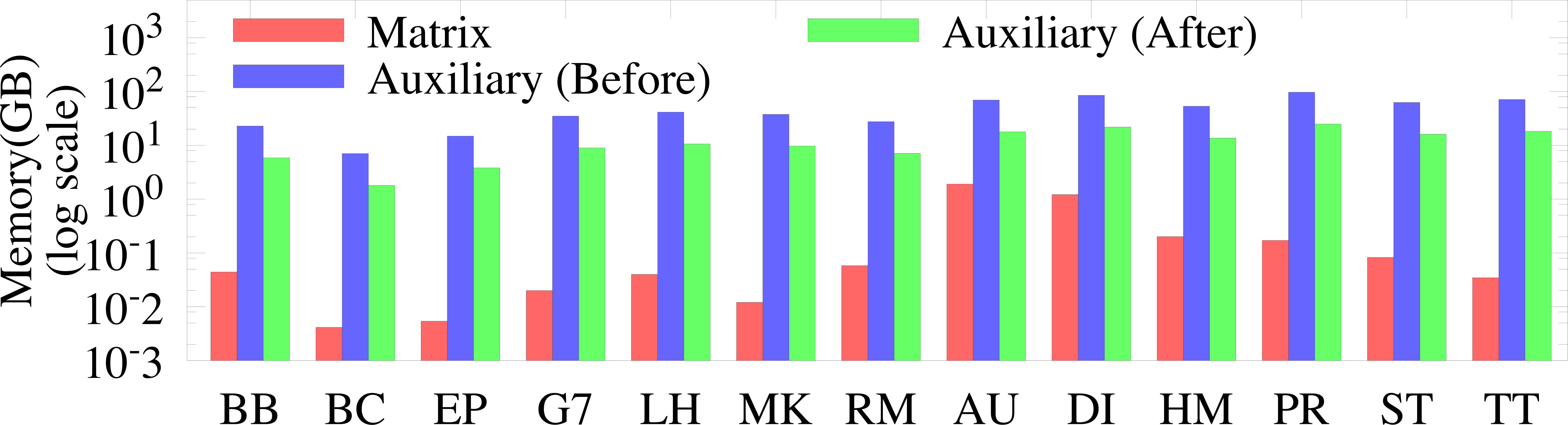} 
\vspace{-.05in}
	\caption{{{\sofa} memory consumption of the original matrix vs the auxiliary data structures on a Summit compute node.}
	\vspace{-.1in}
	}%
	\label{fig:stacked_aux_mat_memory_lines} 
\end{figure}

While the fast runtime is important for symbolic factorization, so is the
small space consumption. As shown in  Figure~\ref{fig:alg_back}(b), the original fill2 algorithm requires three auxiliary arrays, fill(:), frontierQueue(:) and newFrontierQueue(:). When we improve the fill2 algorithm with combined frontier queue traversal, we need to add two new arrays: tracker(:) and newTracker(:). In addition, we need maxId(:) to allow the revisitation mechanism to avoid false negatives. 
Altogether, each source needs six nontrivial auxiliary data structures, each of which consumes memory space of size $\mathcal{O}(|V|)$.
Further, since we often need tens to thousands of concurrent sources to saturate a GPU (discussed in Section~\ref{sec:parallel}), the memory requirement of the initial version of {\sofa}  for certain matrices becomes significantly large in comparison to the available memory on GPUs.  
Figure~\ref{fig:stacked_aux_mat_memory_lines} presents the GPU memory consumption for the sparse matrix and the auxiliary data structures (before and after space optimization) mentioned
in Table~\ref{tab:complexity} on one Summit node. One can observe that the memory requirement for the auxiliary data structures is orders of magnitude larger than that for the matrices. In particular, maximum and minimum memory consumption ratios for auxiliary data structure and matrices are 3121:1 in MK and 36:1 in AU, respectively. 

Section~\ref{sec:space} presents a series of optimizations to reduce the auxiliary data structures memory consumption while maintaining similar performance.
As shown in Figure~\ref{fig:stacked_aux_mat_memory_lines}, these optimizations
reduce the memory consumption of the auxiliary data structure by an average factor of 4$\times$. The maximum and minimum ratios of memory consumption for auxiliary data structure and matrices are reduced to 801:1 in MK and 9:1 in AU.

\begin{figure}[t]
	\centering
	\includegraphics[width=1.05\textwidth]{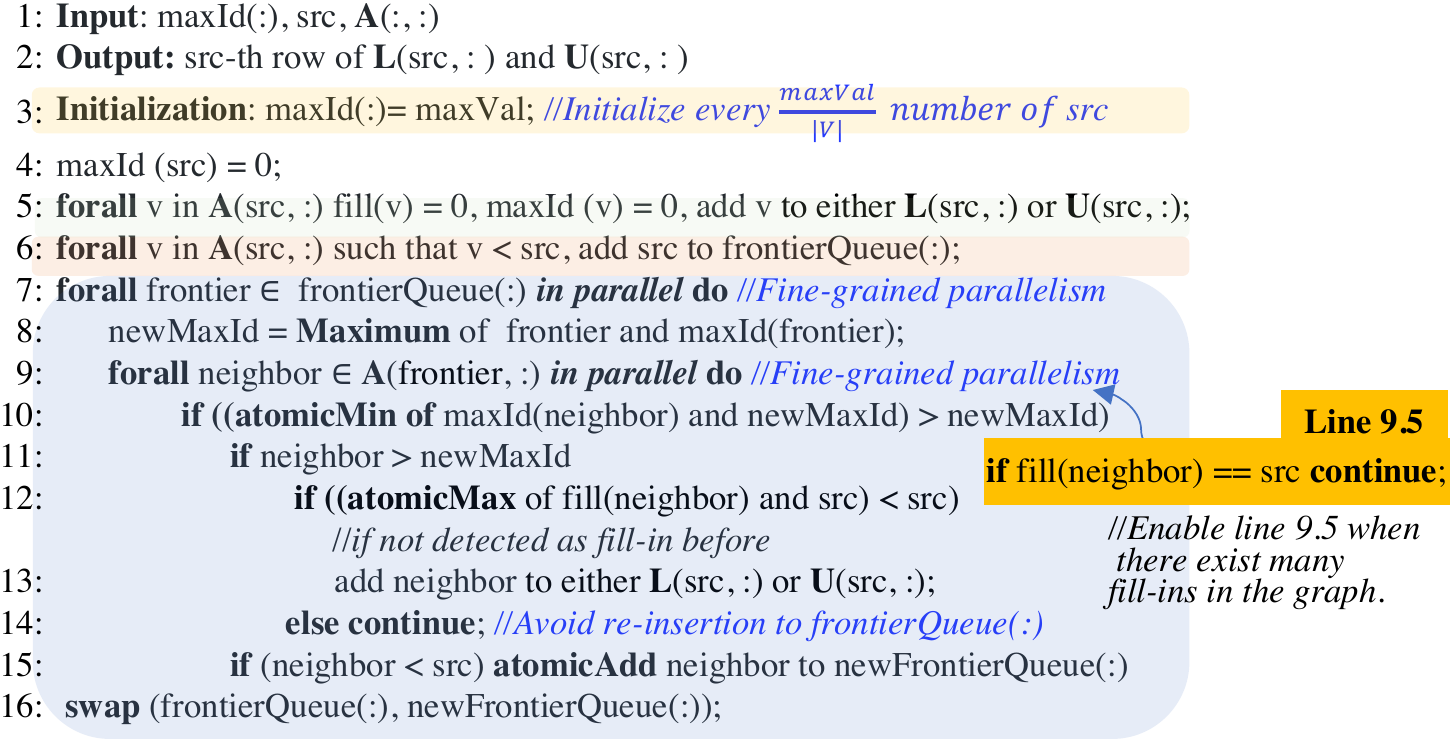} 
	\caption{{\sofa} parallel algorithm. The data structures are in MATLAB format. $\mathbf{L(src, :)}$ and $\mathbf{U(src, :)}$ represent the $\mathbf{L}$ and $\mathbf{U}$ structures of the row \textbf{src}. 
	\vspace{-.15in}
	} 
	\label{fig:alg} 
\end{figure}

\section{A New Fine-Grained Parallel Symbolic Factorization Algorithm}
\label{sec:parallel}



{\sofa} addresses the first challenge, i.e., limited fine-grained parallelism within traversal for a single row,
by \textit{i) allowing parallel processing of the frontiers, (ii) adding a new data structure {maxId(:)} array of size $|V|$, and (iii) allowing revisitation of vertices.} 
Below we explain the data structure and algorithm design separately. 
Formally, from the source vertex $src$ to a specific vertex $v$, there often exist multiple paths, and each path has a maximum numbered vertex. {\sofa} uses {maxId}(v) to store the minimum of all the maximums of the paths from $src$ to $v$.

{\sofa} allows repeated updates to {maxId(:)} along the traversal. And the traversal terminates once the {maxId} values of all the vertices converge to their minimal value. 
As shown in Figure~\ref{fig:alg}, in the beginning, the neighbors of \textit{src} that are smaller than $src$ are eligible for continuing traversal along these neighbors. Hence they are inserted into frontierQueue(:) at line 6. 
During traversal, the algorithm allows all the frontiers to explore the graph in parallel. {With initial setting of 0 for every vertex, fill(v) of a vertex v in the algorithm is set to \textit{src} if there is a nonzero (src, v) in the filled graph. The value of fill(:) helps avoid the re-detection of fill-ins at line 12. For a vertex v, fill(v) $<$ \textit{src} means that vertex v is not yet in the filled structures of row \textit{src}.}

{\sofa} fulfills two tasks in lines 9 - 15. The first task is to check whether a neighbor introduces a fill. The second task is to decide whether a vertex can become a frontier. 
For both tasks, we need the new path from the frontier to the current neighbor to change the {maxId} of this neighbor, that is, maxId(neighbor) should be updated by newMaxId at line 10. The function atomicMin updates the maxId(neighbor) by the minimum of maxId(neighbor) and newMaxId atomically at line 10. For the first task, this neighbor further needs to satisfy Theorem~\ref{thm:fill}. And this neighbor should not have introduced a fill before, i.e., line 12. Otherwise, we face the issue of re-insertion of this neighbor into either $\mathbf{L}$ or $\mathbf{U}$.
{For the second task, a neighbor further needs to meet another criterion in order to become a frontier, that is, this neighbor is smaller than the source (line 15).}
Otherwise, this neighbor either cannot be a frontier.

\begin{figure}[t]
	\centering
	\includegraphics[width=.97\textwidth]{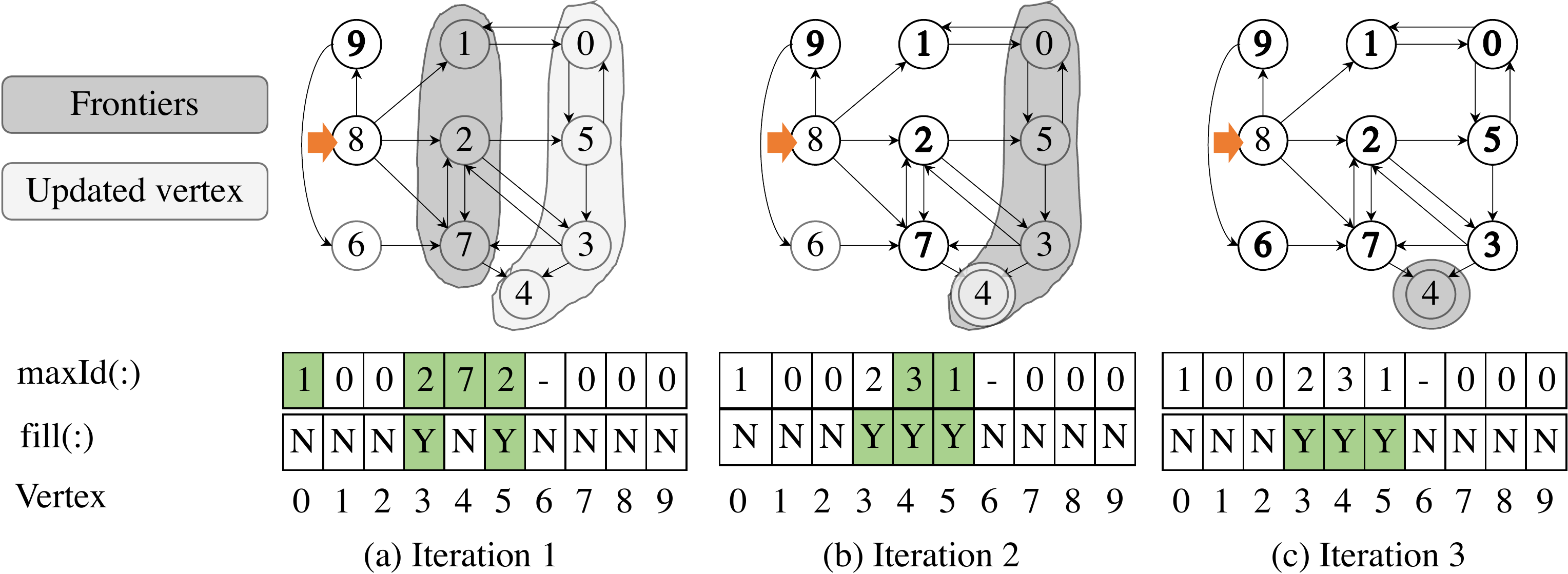} 
	\caption{{\sofa} traversing the graph from Figure~\ref{fig:matrix_begin_end}(b). The  dark, and light gray regions, respectively, represent the current frontier and vertices with maxId(:) updated. Vertex 8 is the source. The maxId(:) and fill(:) are at size of $|V|$. {Note, Y and N in fill(:) are respectively used to show if the corresponding vertex is detected as fill-in or not.}
	\vspace{-.2in}
	}
	 
	\label{fig:fill_in} 
\end{figure}

\textbf{Example}. To aid the understanding of {\sofa}, Figure~\ref{fig:fill_in} demonstrates how {\sofa} traverses from $src = 8$ on {the graph of Figure~\ref{fig:matrix_begin_end}(b)}. 
After line 5, the values in maxId(:) corresponding to the vertices \{1, 2, 7, 8, 9\} become 0, while the rest remain as $|V|$ (i.e., `-' in this case). 


At iteration 1, the frontiers \{1, 2, 7\}, in parallel, traverse their neighbors {\{0\}}, \{3, 5, 7\} and \{2, 4\}, arriving at updated maxId(:) and fill(:) array at the bottom of Figure~\ref{fig:fill_in}(a). Subsequently, we obtain 1, 2, 7 and 2 as the {maxId} along the paths $8\rightarrow 1\rightarrow 0$, $8\rightarrow 2\rightarrow 3$, $8\rightarrow 7\rightarrow 4$ and $8\rightarrow 2\rightarrow 5$. Clearly, these paths introduce fills at 3 and 5 as shown in the fill(:). Proceeding to iteration 2, frontiers \{0, 3, 4, 5\} generate new paths as $8\rightarrow 1\rightarrow 0\rightarrow 5$, $8\rightarrow 2\rightarrow 3\rightarrow 4$  and $8\rightarrow 2\rightarrow 5\rightarrow 3$. Even path $8\rightarrow 1\rightarrow 0\rightarrow 5$ updates maxId(5) at line 10, it will not introduce a fill due to line 12, thus 5 will not be enqueued into newFrontierQueue(:). Path $8\rightarrow 2\rightarrow 5\rightarrow 3$ stops at line 10 because the stored maxId(3) = 2 is smaller than the newMaxId = 5. Finally, path $8\rightarrow 2\rightarrow 3\rightarrow 4$ qualifies line 10 since the stored maxId(4) = 7 is greater than newMaxId = 3, as well as introduces a new fill due to line 12 is true. Afterwards, only 4 is in the newFrontierQueue(:) at iteration 3 which will not introduce any new frontiers. 


It is important to note that the parallel traversal to all the neighbors of a frontier is not possible by simply allowing to process multiple threshold vertices in parallel, as mentioned in Challenge \#1 (lines 9 - 10 of fill2 algorithm). 
Now, we discuss how the detection of the fill-in (8,4) from Figure~\ref{fig:fill_in} may not occur when we allow multiple threads to traverse the neighbors \{1, 2, 7\} in parallel in fill2. 
At line 6 in fill2 algorithm, the fill(:) entries of all the neighbors \{1, 2, 7\} of source 8 are marked as 8. Assuming multiple threads can work in parallel at line 9, implying the thresholds will be 1, 2 and 7 from three different threads, respectively. In that case, all the neighbors 1, 2 and 7 will be inserted into the frontierQueue(:). When accessing the neighbor of the frontier at line 12, if a thread working on frontier 7 explores vertex 4 before any other threads, fill2 algorithm will update fill(4) = 8 without detecting a fill-in at (8, 4) because the path $8\rightarrow 7\rightarrow 4$ does not satisfy Theorem~\ref{thm:fill}. Later, the thread that could detect a fill
at (8, 4) because of path $8\rightarrow 2\rightarrow 3\rightarrow 4$ fails to do so because this thread cannot enter the branch at line 13
since fill(4) is already updated as 8. Hence, the fill-in (8, 4)
remains undetected if we seek fine-grained parallelism in straightforward fill2 algorithm at Figure~\ref{fig:alg_back}(b).

\textbf{Optimizing the initialization of {maxId(:)}.}
Maintaining a {maxId(:)} array for every traversal would require an excessive amount of space that becomes impractical for large datasets considering limited memory space on modern GPUs. Hence reusing maxId(:) array for different sources becomes essential. But this reuse also comes with reinitialization overhead, that is, making maxId(:) to maxVal before traversal. And this overhead is nontrivial, e.g., {it takes 22\% of the total time for PR dataset to re-initialize the {maxId(:)}}. 

To reduce the reinitialization overhead, we propose to divide the value range $[0, maxVal]$ of maxId(:) into smaller ranges, i.e., $[0, |V|]$ range for each source. Note that maxVal is $2^{32}$ for a 32-bit integer type. During traversal, different sources can work on their respective value range of maxId(:) without reinitialization. For instance, for the first source, the traversal updates the {maxId} of the vertices in the range of [maxVal - $|V|$, maxVal]. Moving to the next source, we regard the range of [maxVal - $2\cdot |V|$, maxVal - $|V|$] as valid for {maxId}. In this context, any {maxId} value greater than the upper bound, which is maxVal - $|V|$ in this case, is treated as initialized. Note, the value of maxId will not be smaller than maxVal - $|V|$ before execution. 
 This optimization helps skip the maxId(:) initialization for a total of $\frac{maxVal}{|V|}$ sources. For instance, it helps reduce the ratio of reinitialization over the total time consumption from 22\% to 0.082\% for PR dataset.




\textbf{Optimizing the access to maxId(:) and fill(:)}.
It is important to mention that the accesses to both maxId(:) and fill(:) array are random thus time consuming. One can either access maxId(:) first to reduce the follow-up access to fill(:) or vice versa (i.e., adding line 9.5 to the pseudo-code). 
In both cases, we {avoid repeated frontier enqueuing} for vertices that are already detected as fill-ins at line 14. 

Putting maxId(:) access before fill(:), which is the pseudo-code in Figure~\ref{fig:alg}, will avoid the access to fill(:) array when the new path fails to update the {maxId} of existing paths, i.e., line 10 is evaluated as false. Consequently, this path avoids the access to follow-up fill(:) at line 12. Path $8\rightarrow7\rightarrow4$ from Figure~\ref{fig:fill_in}(a) falls in this case. 

Adding line 9.5 to the pseudo-code in Figure~\ref{fig:alg} will avoid unnecessarily lowering the {maxId} of an {already detected fill-in}. Note, we do not need to do so because this vertex will always propose its own vertex ID as the maxId(:) for the paths that come across this vertex. 
Using Figure~\ref{fig:fill_in}(a) as an example, this logic avoids lowering {maxId}[5] from 2 to 1 when the path $8\rightarrow1\rightarrow0\rightarrow5$ attempts to do so because (8, 5) is already a fill due to  $8\rightarrow2\rightarrow5$. And the continuation of the paths from 8 through 5 will surely use 5 as the {maxId}. 

Including line 9.5 or not is a graph dependent option. Particularly, adding line 9.5 is beneficial for graphs that have relatively larger number of fill-ins. But for graphs with relatively smaller number of fill-ins, the condition of line 9.5 will become false for most of the time, resulting in higher overhead than benefits. 



\textbf{Multi-source symbolic factorization with combined frontierQueue(:)}  executes multiple sources of the Algorithm in Figure~\ref{fig:alg} concurrently on a single GPU in order to saturate the computing resources. To balance the workload across threads, we combine the frontiers of various concurrent traversals into a single frontierQueue(:), which is analogous to recent multi-source graph traversal projects~\cite{liu2016ibfs}. We further make three revisions to this design. First, we need to use maxId(:) instead of a single bit to track the status of each vertex. Furthermore, {\sofa} relies upon tracker(:) and newtracker(:) to identify the source of each frontier.
Last but not least, we directly use atomic operations to enqueue frontiers into the combined frontier queue, which is faster than prefix-sum based approach, according to~\cite{gaihre2019xbfs}.

\section{Parallel Efficient Supernode Detection} 
\label{sec:supernode}

This section further introduces a \textit{massively parallel supernode detection algorithm} which can \textit{balance the workload}, \textit{minimize the communication} and \textit{saturate the GPU resources} with judicious source scheduling and unified memory assistance. 

\textbf{GPU-friendly supernode detection with shared memory optimization.}
We introduce a two-phase supernode detection design to match the SIMT nature of GPUs. First, {we use one GPU thread to examine whether the nonzero count of a row in $\mathbf{U}$ is one fewer than the previous row (i.e., requirement i), and use a bitmap in shared memory to keep track of this information. If requirement (i) is not met, it means the current row starts a new supernode. We will use a queue in shared memory to keep track of this leading row.} 
Second, each GPU thread dequeues a leading row of a supernode, examines the bitmap, also assesses the requirement (ii) of a supernode in Definition~\ref{def:T3}. Only when both requirements are met, the current row is considered to be part of this supernode. \textit{The novelty of this two-phase design is that the first phase helps break the chunkSize of rows into independent subranges so that our second phase can work on these subranges in parallel, which fits the SIMT nature of GPUs.} In addition, the first phase is also massively parallel.

\begin{figure}[h]
	\floatbox[{\capbeside\thisfloatsetup{capbesideposition={right,top},capbesidewidth=3.2cm}}]{figure}[\FBwidth]
	{
		\caption{Parallel supernode detection for matrix $\mathbf{U}$ in Figure~\ref{fig:matrix_symbolic}(c), where Phase I confirms $nnz(\mathbf{U}(s, :)) = nnz(\mathbf{U}(s-1,:))-1$ while Phase II checks $\mathbf{L}(s, r) \neq 0$.
		\vspace{-.1in}
			}
	\label{fig:supernode_parallel}
	}
	{
		\includegraphics[width=.95\linewidth]{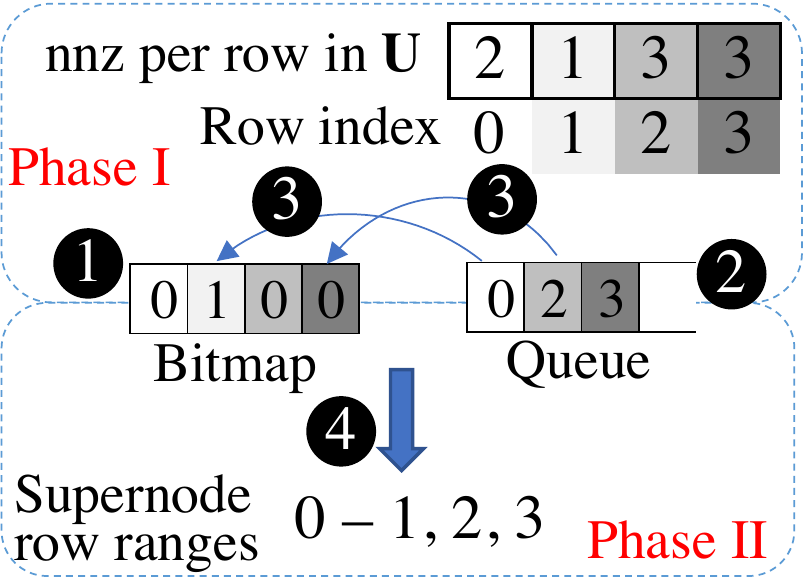}
	}
\end{figure}

Figure~\ref{fig:supernode_parallel} demonstrates how to, in parallel, detect supernodes for row ranges 0 - 3 of matrix $\mathbf{U}$. In phase I, all threads in parallel check whether the nnz of current row is 1 fewer than the prior row. Since only row 1 satisfies the condition, the bitmap is set to ``0100'' (\circled{1}). Further, the leading rows of supernode are \{0, 2, 3\} in queue (\circled{2}). In phase II, each thread is assigned to one leading row entry, and checks whether the bit following the leading row is 1 (\circled{3}), as well as the corresponding entry in $\mathbf{L}(s, r) \neq 0$ (\circled{4}). If both conditions are met, the supernode grows. We will continue this process until no supernode grows. Since row 1 satisfies both conditions while row 2 does not, supernode of row 0 grows to include only row 1.

\textbf{Communication- and saturation- aware inter-node source scheduling.}
Since inter-node communication is significantly more expensive than intra-node counterpart, we restrict the supernode detection inside of each computing node. We use the following equation to perform inter-node source scheduling:
\begin{equation}\label{eq:source}
\begin{split}
    &chunkSize\times numConcurrChunksPerNode\\
    &= \#C \times numGPUsPerNode,
\end{split}
\end{equation}
where $chunkSize$ is identical to the size of the user defined maximum supernode. Note, we want the size of a chunk to be neither a fraction nor multiple of the maximum supernode size. On the one hand, since we interleave consecutive chunks to different nodes, making $chunkSize$ a fraction of the maximum supernode size will introduce inter-node communication during supernode detection. On the other hand, making $chunkSize$ a multiple of supernode size will make chunk size too large, potentially lead to worse workload imbalance. Further, $\#C$ is the preferred number of sources to saturate one GPU. In short,  all the three items in Equation~(\ref{eq:source}), i.e., $chunkSize$, $\#C$, and $numGPUsPerNode$ are known, one can derive the value for $numConcurrChunksPerNode$. For example, the PR dataset on one Summit node would need 6,144 sources to saturate the GPUs. If the defined supernode size is 128, \textit{numConcurrChunksPerNode} would be 48. 

\textbf{Unified memory assisted intra-node source scheduling.} {Once the number of concurrent chunks for a computing node is decided, we assign the sources of these chunks to various GPUs of this node in an interleaved fashion in order to balance the workload.}
However, since supernode detection requires the nonzero count of consecutive rows, communicating the fill information between GPUs is needed. 
Specifically, for a supernode spanning from row $r$ to $s-1$, one needs to communicate the \textit{nonzero count} of row $s$ in $\mathbf{U}$, and whether there is a \textit{nonzero at $\mathbf{L}(s, r)$}. Since both data sizes to communicate are small and we cannot predict which nonzero count and fill location are required beforehand, traditional \textit{cudaMemcpy} based data transfer is not suitable for such a kind of data sharing. 

To this end, we choose the unified memory option over explicit data transfer. In this context, {\sofa} uses a single CPU process to manage all the GPUs in each node, where the kernel launches are performed asynchronously. It is also important to note that unified memory introduces several configurations which concern the performance. 

First, we choose remote access over page migration when accessing the unified memory. Since we interleave the sources assignment in a fine-grained manner, various GPUs might compete to migrate the same page that stores the fill(:) or nonzero count information from the source GPU. This will hurt the performance. During implementation, our key guideline is \textit{keeping this data in the GPU that modifies this data}. Particularly, only one GPU will detect the fill information for a specific row (i.e., modify that fill(:)), so does that for the supernode detection. This implies that only one GPU needs to keep the fill(:) information of that row locally while the remaining GPUs will access that information remotely. In implementation, we use the \textit{cudaMemAdviseSetPreferredLocation} flag to {advise the unified memory driver} to keep that data in the same GPU which modifies this data~\cite{cudaMemAdvise}.
For the remaining GPUs, we use \textit{cudaMemAdviseSetAccessedBy} flag to instruct them to access that data remotely.



Second, {\sofa} allocates separate fill(:) memory for different GPUs instead of using one virtual space spanning across all GPUs in one node. This design is inspired by a key observation that there exist a significant number of page faults when all the GPUs share one fill(:) address. This is caused by the fact that this single memory space is not perfectly aligned from one GPU to another. That is, one page could span across two GPUs. In this case, once both GPUs need to modify that page, these GPUs will compete for that shared page for writing. {This will lead to frequent page migration.}

\section{Space Optimization} \label{sec:space}




Continuing our discussion in Challenge \#3 of Section~\ref{sec:challenge}, this section will rigorously quantify the \textit{space crisis faced by multi-source concurrent symbolic factorization}. 
Table~\ref{tab:complexity} presents the space complexity of the six major data structures used by {\sofa}, namely, \textit{frontierQueue(:)} \& \textit{newFrontierQueue(:)}, \textit{tracker(:)} \& \textit{newtracker(:)}, \textit{maxId(:)}, and \textit{fill(:)}. When the number of concurrent sources is \#C, the total space consumption would be around $6\cdot|V|\cdot\#C$ entries. Apparently, space consumption immediately becomes a key problem for large graphs with relatively big number of concurrent sources. 
However, when it comes to performance, {\sofa} prefers a larger \#C which will provide more workload to better saturate the GPU computing resources. 


\begin{table}[hbt!]
{
\vspace{-.05in}
\scriptsize
\begin{center}
\begin{tabular}{|c|c|}
\hline
\textbf{Data structure} & \textbf{Space complexity} \\ \hline
\begin{tabular}[c]{@{}c@{}}frontierQueue{(:}{)} \& newFrontierQueue{(:}{)}\end{tabular} & $2\cdot |V|\cdot \#C$\\ \hline
\begin{tabular}[c]{@{}c@{}}tracker{(:}{)} \& newTracker{(:}{)}\end{tabular} & $2\cdot |V|\cdot \#C$\\ \hline
maxId{(:}{)} & $|V|\cdot \#C$ \\ \hline
fill{(:}{)}  & $|V|\cdot \#C$\\ \hline
\end{tabular}
\end{center}
}
\vspace{-.15in}
\caption{The data structures used by multi-source concurrent {\sofa}, where $|V|$ and \#C are the number of vertices in the graph, and concurrent sources, respectively.
\vspace{-.1in}
}
\label{tab:complexity}
\end{table}

\begin{table}[hbt!]
\begin{center}
{\scriptsize
\begin{tabular}{|c|c|c|}
\hline
\textbf{Dataset} & \textbf{Average usage (\%)} & \textbf{Peak usage (\%)} \\ \hline
AU & 0.12  & 8.2    \\ \hline 
DI & 0.04  & 4.55  \\ \hline
HM & 0.01  & 1.7    \\ \hline 
PR& 0.11 & 25.0           \\ \hline
ST & 0.1 & 1.0           \\ \hline
TT & 0.1 & 11.0           \\ \hline
\end{tabular}
}
\end{center}
\vspace{-.1in}
\caption{{Percentage of usage of allocated frontier queues.
\vspace{-.1in}
}
}
\label{tab:FQ_usage}
\end{table}


In this section, we propose three interesting optimizations based upon the access pattern and usage of various data structures to combat the high space complexity. Particularly, we introduce external GPU frontier management, bubble removal in maxId(:), dynamic space allocation to dynamically assign memory space to various data structures. Note, this dynamic space allocation also allows {\sofa} to support configurable space consumption.




\textbf{External GPU frontier management} is motivated by the observation in Table~\ref{tab:FQ_usage} where the average usage of frontier-related data structures remains
low for the large datasets, i.e., AU, DI, HM, ST and TT. 
{However,} the peak usage can rise to as high as 25\% for PR. This observation implies that we can allocate relatively smaller memory space to hold frontier-related data structures because the usage remains low for vast majority of the iterations. Once the usage goes beyond the allocated space, we resort to our external-GPU option which is presented below.



	

\begin{figure}[h]
	\centering
	\includegraphics[width=.9\textwidth]{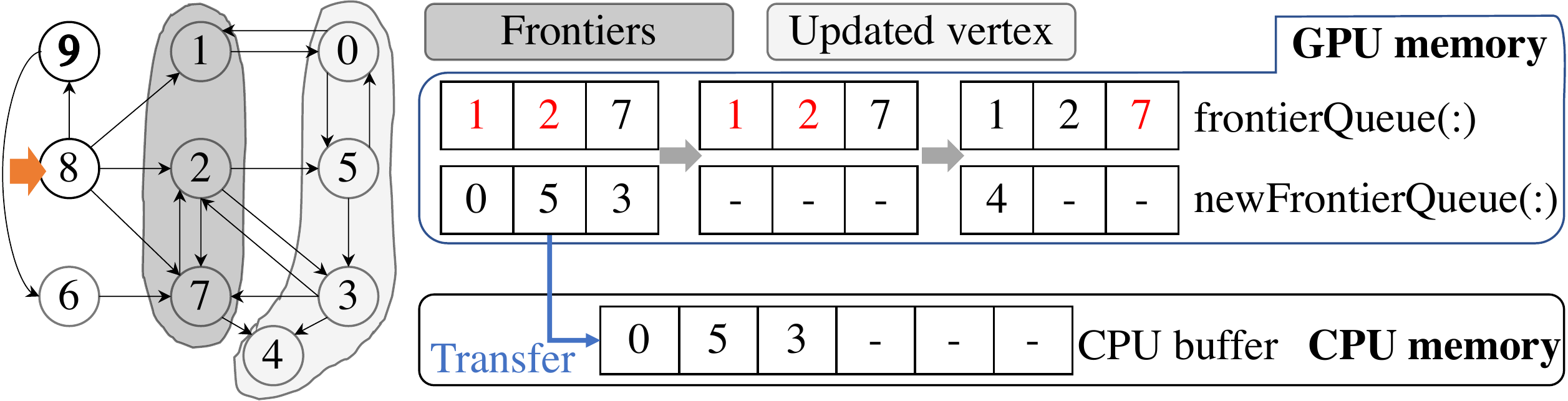} 
	\caption{External GPU frontier management. {Note, the traversal iteration is same as in Figure~\ref{fig:fill_in}(a).}
	  \vspace{-.1in}
	}
	\label{fig:space_complexity_optimization}
\end{figure}

We propose to only allocate a fraction of the required space on GPU for the four frontier-related data structures, i.e., frontierQueue(:), newFrontierQueue(:), tracker(:) and newtracker(:) and write the extra frontiers out of GPU. Then, we load these external GPU data structures in GPU for computation.  Figure~\ref{fig:space_complexity_optimization} demonstrates this design.
For brevity, Figure~\ref{fig:space_complexity_optimization} only uses a single thread to traverse the graph
for source 8 without loss of generality. The size of the allocated newFrontierQueue(:) is only three. At iteration 1, frontiers 1 and 2 exhaust the newFrontierQueue(:) by their neighbors 0, 5 and 3. Subsequently, we copy these three frontiers to CPU memory to still have available space for the incoming frontiers, 7 in this case. {Proceeding to the next iteration that comes after swapping the queues, we will finish frontier 7 in the frontierQueue(:) first and load the frontiers from CPU in GPU for further computation.}
It is worthy of mentioning that, instead of directly storing the source ID in tracker(:) and newtracker(:) in multi-source concurrent traversal, we propose to store the index of the source for each frontier to reduce the space further.



\textbf{Bubble removal in {maxId(:)}} is supported by the key observation that a source vertex is not allowed to traverse vertices that are larger than the source. Consequently, we can remove the ``bubbles'' in the {maxId(:)} that are larger than the source. Here, ``bubble'' means the allocated space that is not used in maxId(:).
Assuming the source vertex is $v$, according to Theorem~\ref{thm:fill}, we will never update the {maxId} of the vertices that are larger than $v$. 


\textbf{Dynamic space allocation across data structures} is motivated by two facts. First, maxId(:) and fill(:) accesses are more random than frontier-related data structure. Particularly, the accesses to maxId(:) and fill(:) are determined by the frontier's neighbors which often have random vertex IDs. Therefore, {\sofa} needs to put the entire maxId(:) and fill(:) arrays in GPU memory in order to achieve desirable performance. Second, the space requirement maxId(:) is dynamic, that is, smaller sources need smaller maxId(:) and vice versa. Therefore, we can dynamically adjust the maxId(:) space so that the frontier-related data structures can have more GPU resources when possible.

Towards this end, 
we first allocate a big chunk of memory, in contrast to allocate separate memory spaces for various data structures discussed in Table~\ref{tab:complexity}. Subsequently, this memory chunk is dynamically divided among the data structures with priority given to maxId(:) and fill(:).
For a given number of concurrent sources in a traversal, after the space reduction strategy of {maxId(:)}, the amount of memory required for maxId(:) remains low for smaller sources. In this scenario, we can allocate more space for other data structures. 
Once we start working on larger sources, the maxId(:) space requirements begin to climb.
In this context, we prioritize the space requirement for maxId(:) {along with fill(:)} so that a large number of concurrent sources can execute together {with better, at least sustained, performance}.



\textbf{{\sofa} with configurable space budget.} After putting all the major data structures of {\sofa} into one continuous memory space, we enable a new feature, i.e., configurable memory budget for {\sofa}. This optimization will first conduct bubble removal, {dynamic space allocation} and external GPU frontier-related data structure management. If {\sofa} still suffers from space shortage, {\sofa} will judiciously reduce the number of concurrent sources in order to restrict the memory space consumption in the given budget. 

\section{Evaluation} \label{sec:eval}



\begin{figure}[h]
	\centering
	\includegraphics[width=.95\textwidth]{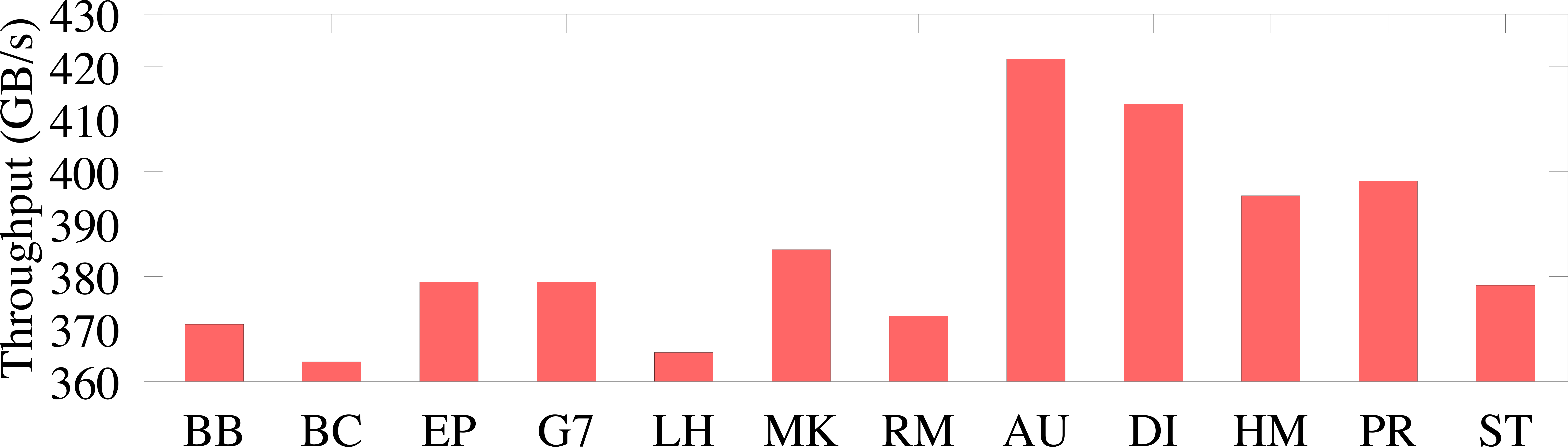} 
	\caption{Throughput achieved by {\sofa} on a V100 GPU.
	}

	\label{fig:throughput_1_GPU} 
\end{figure}

\begin{figure}[h]
	\centering
	\includegraphics[width=.95\textwidth]{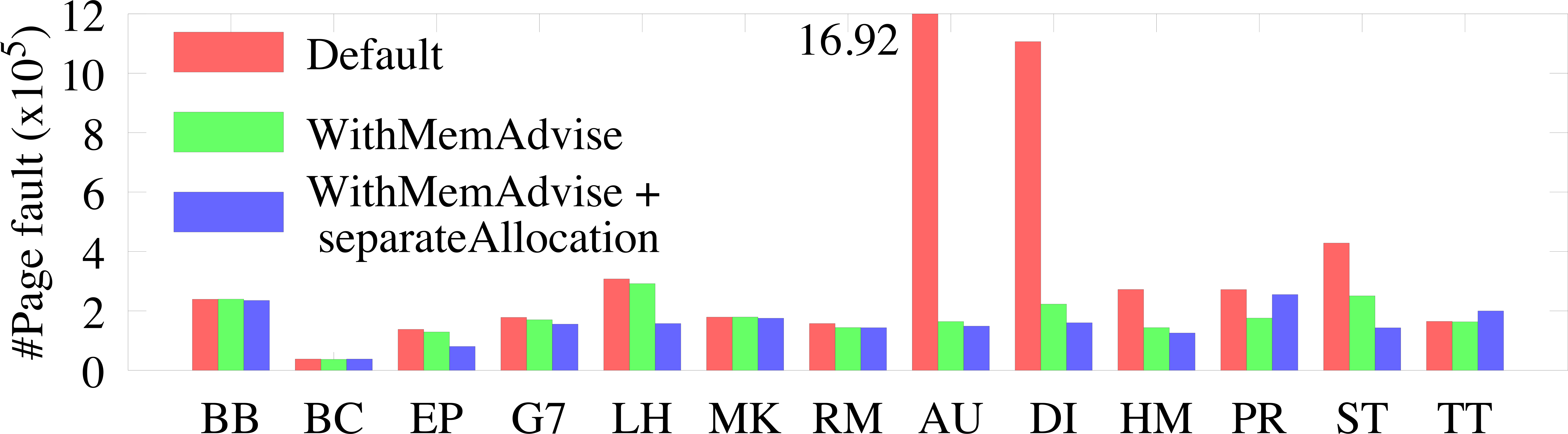} 
	\vspace{-.1in}
	\caption{Performance impacts of unified memory optimization on one Summit node. 
	  \vspace{-.15in}
	} 
	\label{fig:page_fault} 
\end{figure}

We implement {\sofa} with {$\sim$2,500} lines of C++/CUDA code, and compile the source code with NVIDIA CUDA 10.1 Toolkit with the optimization flag set to be O3. We use IBM Spectrum MPI 10.3.0.0 for inter-node communication. The evaluation platform is the Summit supercomputer at Oak Ridge National Laboratory~\cite{summit}, where each computing node is equipped with {dual-socket IBM POWER9 CPU processors (i.e., {42} cores)}, and six NVIDIA V100 GPUs. All the GPUs on one Summit node are connected with NVIDIA's high-speed NVLink.
We use Traversed Edges Per Second (TEPS) to report the graph traversal performance, and take the average of three runs. We use NVIDIA nvprof profiling tool for Figures~\ref{fig:throughput_1_GPU} and~\ref{fig:page_fault}.

\begin{figure}[hbt!]
	\centering
	\includegraphics[width=.95\textwidth]{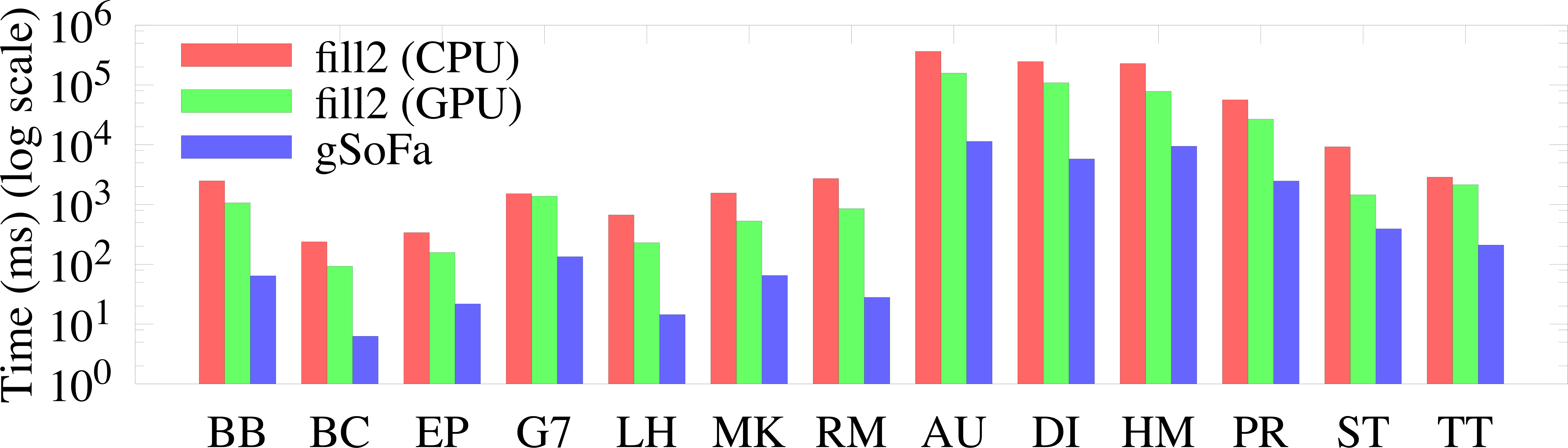} 
	\caption{Performance of the parallel CPU and GPU versions of fill2 and {\sofa} on a Summit compute node.
	\vspace{-.1in}
	}

	\label{fig:speedup_Fill2} 
	  \vspace{-.1in}	
\end{figure}

Figure~\ref{fig:speedup_Fill2} compares the original fill2 algorithms on CPU and GPU, and {\sofa} on one Summit node. Both CPU and GPU versions of fill2 that we are using are the straightforward parallel implementation of the fill2 algorithm in Figure~\ref{fig:alg_back}(b). The parallel CPU version employs all the 42 cores in a Summit node to perform symbolic factorization for 42 sources in parallel. The GPU version allows every allocated thread to work in a source as long as the GPU memory is sufficient. On average, {\sofa} is 13.01$\times$ faster than the GPU-based parallel fill2 algorithm, with the maximum and minimum speedups 30.41$\times$ (RM) and 3.70$\times$ (ST), respectively. 
When {\sofa} is compared to CPU version, {\sofa} enjoys even higher speedups, that is, 33.06$\times$, on average, with 97.11$\times$ (RM) and 11.32$\times$ (G7) as the maximum and minimum speedups. This suggests that the original fill2 algorithm is not suitable for massively parallel GPUs. 

\textbf{Comparison with the state-of-the-art CPU algorithms.} 
We compare {\sofa} with the state-of-the-art sequential symbolic factorization in GLU3.0~\cite{peng2020glu3} and the parallel symbolic factorization from SuperLU\_DIST~\cite{grigori2007parallel}. {Note, both the GLU3.0 and SuperLU\_DIST may use GPUs only during numerical factorization or later phases. However, both libraries use CPUs for symbolic factorization.}

\begin{figure}[hbt!]
	\centering
	\includegraphics[width=.95\textwidth]{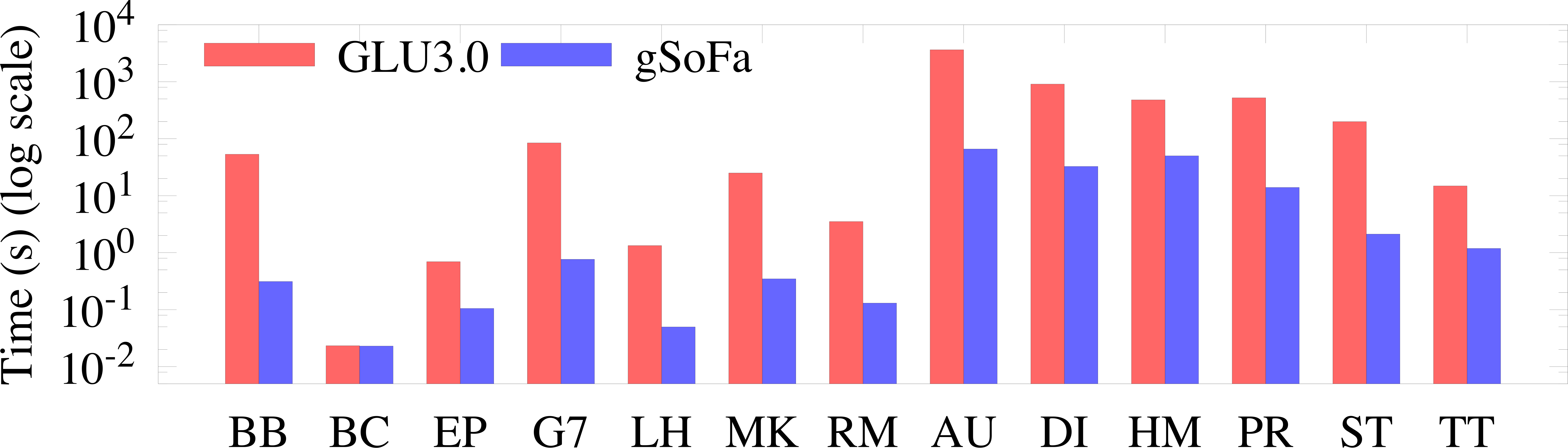} 
	\caption{Performance comparison of {\sofa} with the state-of-the-art CPU symbolic factorization in GLU3.0. For fair comparison, {\sofa} uses one V100 GPU since GLU3.0 performs symbolic factorization sequentially.
	\vspace{-.05in}
	}

	\label{fig:pathbasedVsGLU3} 
	  \vspace{-.1in}	
\end{figure}

 
 Figure~\ref{fig:pathbasedVsGLU3} demonstrates the time comparison of {\sofa} with the symbolic factorization on GLU3.0~\cite{peng2020glu3}. {We limit the {\sofa} to a single GPU because GLU3.0 can only perform single-threaded symbolic factorization on CPU}. Note, GLU3.0 is based upon Gilbert-Peierls algorithm~\cite{gilbert1988sparse} which suffers from stringent data dependency problems if intended to implement in parallel (discussed in Section~\ref{sec:background}). 
In the figure, we can observe a maximum and minimum speedup of 171.1$\times$ (BB) and 1.01$\times$ (BC). {The workload in BC dataset is too small to saturate the GPUs, i.e., see Figure}~\ref{fig:less_workload}(a).
%
%
%
 On average, the {\sofa} is 50.1$\times$ faster than symbolic factorization in GLU3.0.


 \begin{figure}[hbt!]
	\centering
	\includegraphics[width=.95\textwidth]{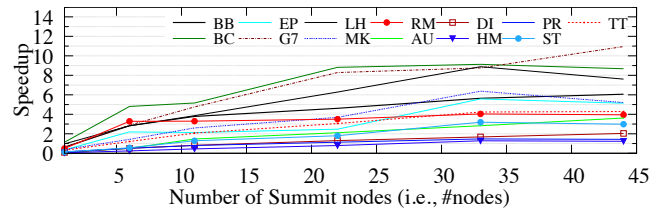} 
	\caption{{Speedup of {\sofa} over the state-of-the-art CPU parallel symbolic algorithm in SuperLU\_DIST.}
	}
	\label{fig:pathbasedVssuperlu} 
\end{figure}

Figure~\ref{fig:pathbasedVssuperlu} compares the performance between {\sofa} and the CPU parallel symbolic factorization in SuperLU\_DIST. {\sofa} starts worse than the CPU parallel algorithm for majority of the datasets. However, with more and more Summit nodes, {\sofa}  begins to outperform the CPU algorithm. Particularly, initially {on one node}, {\sofa} is {1.2$\times$} faster on {BC} and { 1.6$\times$,  2.8$\times$, 1.6$\times$, 1.03$\times$, 2.6$\times$,   2.0$\times$, 13.8$\times$, 11.3$\times$, 19.3$\times$, 6.7$\times$, 7.4$\times$ and 3.2$\times$} slower on BB, EP, G7, LH, MK, RM, AU, DI, HM, PR, ST and TT,
respectively on one node. When it goes to six Summit nodes, {\sofa} bests CPU parallel symbolic factorization on majority of the datasets, i.e., BC, EP, BB, RM, G7, MK, LH and TT. {\sofa} finally outperforms CPU parallel symbolic factorization across all the datasets by 5$\times$, on average, with 44 Summit nodes, at which the maximum speedup of 10.9$\times$ is achieved for G7 matrix and minimum of 1.3$\times$ is achieved for HM matrix.
{For G7, the speedup of {\sofa} is high because G7 is highly non-symmetric which limits the benefits of symmetric pruning in SuperLU\_DIST and its larger workload (i.e., $\frac{\mathrm{nnz(A)}}{\mathrm{Order(A)}}$ in Table}~\ref{tab:datasets}). {HM though has relatively low symmetry but it presents the lowest workload which causes the speedup to be poor.}

\textbf{Throughput.}
Figure~\ref{fig:throughput_1_GPU} demonstrates the throughput achieved by the {\sofa} on V100 GPU. Particularly, {\sofa} achieves a maximum of 421.5 GB/s (47\% of peak memory throughput) for AU dataset and a minimum of 363.8 GB/s (40\% of peak memory throughput). On average, {\sofa} achieves a throughput of 385.2 GB/s (43\% of peak memory throughput) over all the dataset, which is rarely observed for graph traversal related applications~\cite{wang2016gunrock}. 

\begin{figure}[h]
	\centering
	\includegraphics[width=.98\textwidth]{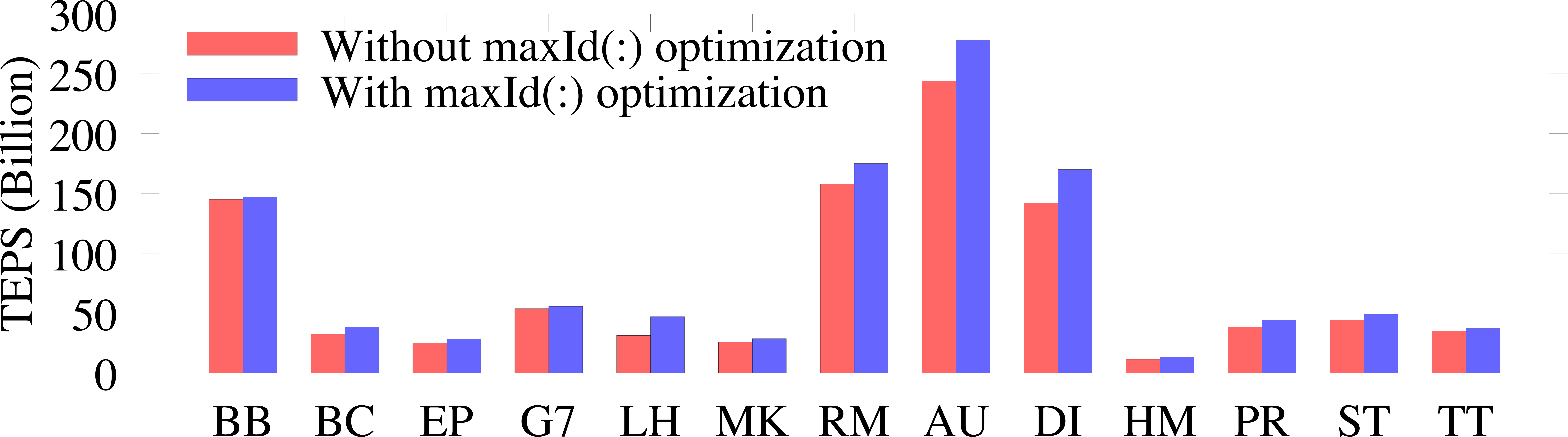} 
	\vspace{-.075in}
	\caption{{Impacts of maxId(:) initialization optimization.
	\vspace{-.05in}
	}
	} 
	\label{fig:max_id_init} 
\end{figure}

\textbf{Performance impact of maxId(:) optimization.} This reduces the initialization overhead for maxId(:). We observe noticeable performance gains across all datasets. On average, we achieve 14\% speedup with this optimization, where the maximum impact comes from LH of 50\% and a minimum of 2\% increment in {BB} dataset.

\begin{figure}[h]
	\centering
	\includegraphics[width=.95\textwidth]{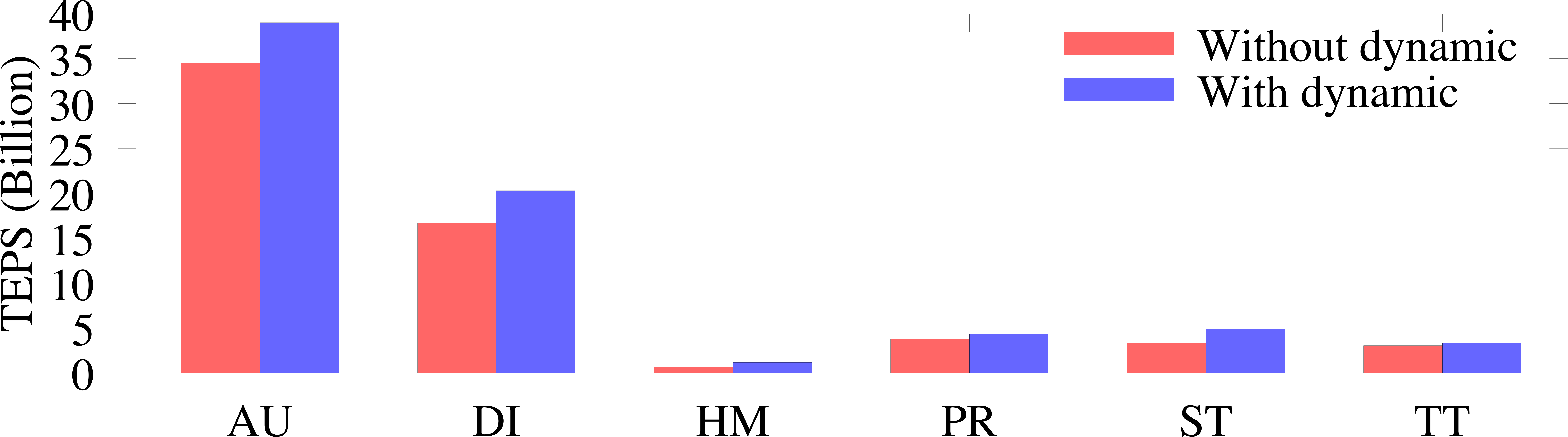} 
\vspace{-.075in}
	\caption{Performance impact of dynamic space allocation on one GPU with 1 GB space allocation. 
	\vspace{-.1in}
	} 
	\label{fig:dynamic_max} 
\end{figure}

\textbf{Performance impact of dynamic space allocation.} Figure~\ref{fig:dynamic_max} presents the effect of dynamic memory allocation. We allocate 1 GB of memory and study the performance of \textit{with} versus \textit{without} the dynamic memory optimization. As expected, this optimization yields performance gains for all the large datasets in Table~\ref{tab:datasets}. 
{Particularly, we observe improvements of {1.2$\times$, 1.2$\times$,}  1.7$\times$, 1.2$\times$, 1.5$\times$ and 1.1$\times$ on {AU, DI,} HM, PR, ST and TT respectively.}





\begin{figure}[h]
	\centering
	\includegraphics[width=.96\textwidth]{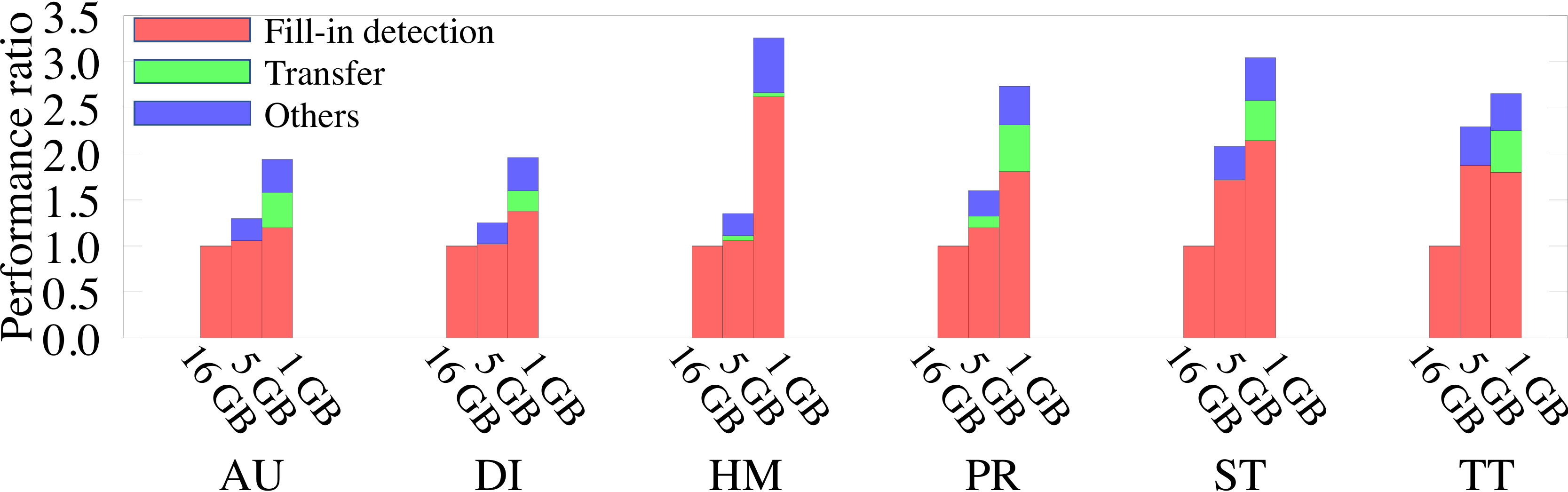} 
\vspace{-.075in}
\caption{Performance study of {\sofa} with different memory capacities on 44 Summit nodes. The allowed GPU memory for all the data structures in {\sofa} varies from {1 to 16 GB}. 
\vspace{-.1in}
}
	\label{fig:diff_ratio_bar} 
\end{figure}

\textbf{{\sofa} space optimization with limited memory budget.} This includes all the space optimizations, further along with an explicit restriction on the available memory space. Particularly, we enable external GPU frontierQueue(:), newFrontierQueue(:), tracker(:) and newtracker(:) management, bubble removal and dynamic allocation. For the single large array that is shared by all the data structures, we limit its size to be 16, 5 and 1 GBs to demonstrate the performance robustness of {\sofa}. This optimization will involve transferring data between CPU and GPU memories and other overheads, such as, checking the condition of memory overflow.

Figure~\ref{fig:diff_ratio_bar} shows the trade-off between the runtime and the space consumption. The general trend is that the performance drops with the decrease of allocated space. Particularly, with merely 1 GB memory budget, the {\sofa} performance decreases by 1.9$\times$, 1.9$\times$, 3.3$\times$, 2.7$\times$, 3.0$\times$ and 2.6$\times$, respectively, on AU, DI, HM, PR, ST and TT datasets. The performance drops in fill-in detection are caused by the fact that a limited memory budget leads to the reduction of \#C. Note, enabling external GPU {\sofa} also results in the overhead of checking whether the frontier queue overflows, which is denoted as ``{\sofa}: Others'' in Figure~\ref{fig:diff_ratio_bar}. 

\begin{figure}[hbt!]
	\centering
	\includegraphics[width=.95\textwidth]{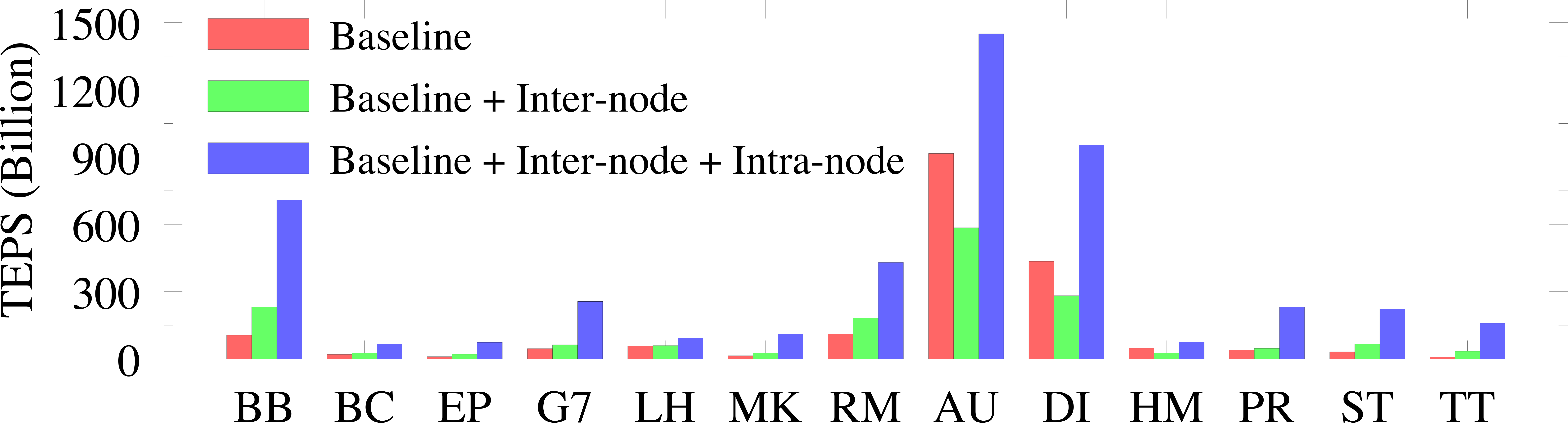} 
	\vspace{-.075in}
	\caption{Performance impacts of supernode detection optimizations on six nodes (i.e., 36 GPUs). Here, the inter-node optimization interleaves the chunk of rows among the compute nodes and the intra-node optimization interleaves rows scheduling in fine-grained manner during fill-in detection to obtain optimum intra-node workload balancing.
	  \vspace{-.1in}
	} 
	\label{fig:different_optimization} 
\end{figure}


\textbf{{Performance impact of supernode detection optimizations.}} In Figure~\ref{fig:different_optimization} we use six nodes (i.e., 36 GPUs) in order to showcase the impacts of our inter-node workload balancing strategies. Specifically, the ``baseline" version assigns a block of continuous chunks of sources to a node. The ``inter-node" interleaves the chunks of sources across different nodes in a round-robin approach. And ``intra-node" performs unified memory-assisted fine-grained source scheduling across GPUs in a node. On average, the inter-node scheduling optimization yields 1.6$\times$ speedup over the baseline. The intra-node optimization further adds another 3.3$\times$ speedup. The maximum gains of inter- and intra- node optimizations are 4.3$\times$ (TT) and 4.9$\times$ (PR), respectively. We also observe performance drops for the inter-node optimization on AU, DI and HM graphs because {the new combination of chunks of sources might have relatively more non-uniform workload distribution}. 

\textbf{Performance impacts of unified memory optimizations.}
Figure~\ref{fig:page_fault} demonstrates the effects of cudaMemAdvise and data structure separation optimizations. We can observe a general trend of reduction in the number of page faults with each optimization's addition. Particularly, cudaMemAdvise reduces the page fault by {2.2$\times$}.
In particular, we see the maximum drop of page faults in AU by a factor of {10.3$\times$}. {The data structure separation optimization, {when added to the cudaMemAdvise version}, further reduces the page faults by an average of {20\%}, with the maximum drop to be {85\%} in LH.} Note, PR and TT experience more page faults due to the data structure separation optimization. This is potentially caused by the fact that the change of memory alignment could result in more page faults for the memory that is not at the boundary.

 \begin{figure}[hbt!]
	\centering
	\includegraphics[width=.9\textwidth]{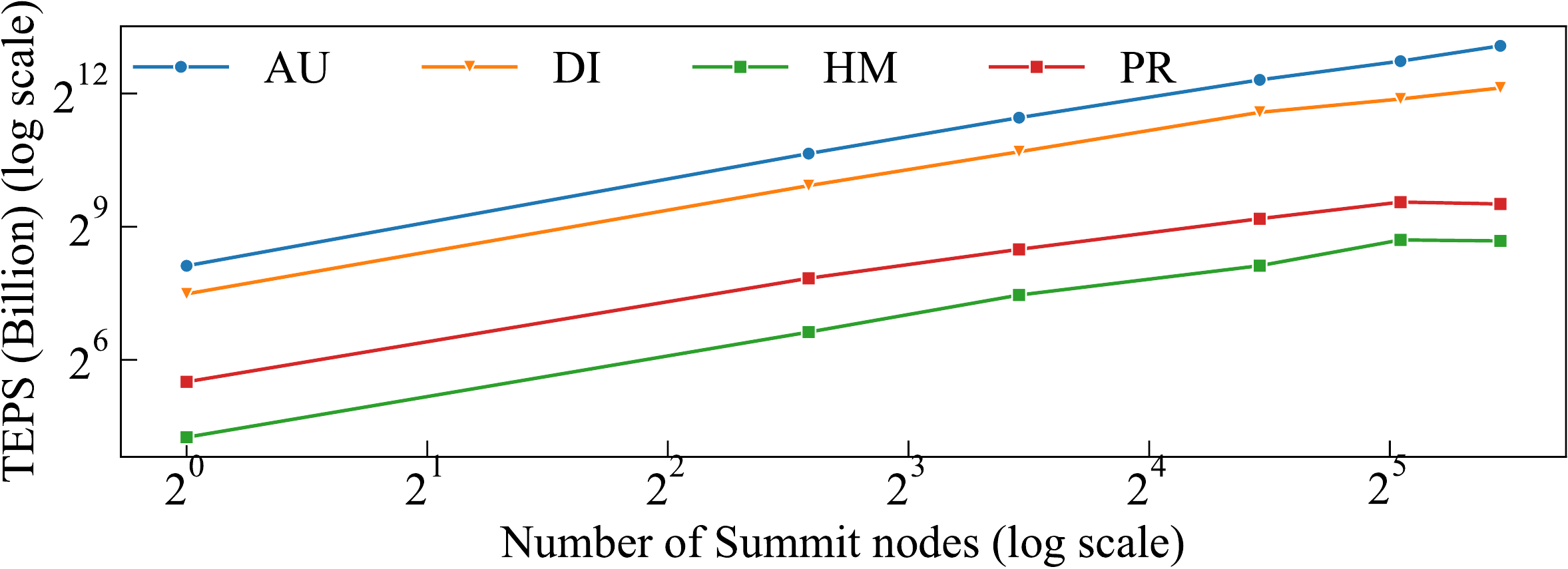} 
	\vspace{-.075in}
	\caption{Scaling {\sofa} to 44 Summit nodes (264 GPUs).
	\vspace{-.075in}	
	} 
	\label{fig:scaling} 
\end{figure}

\textbf{Strong scalability.} {Figure~\ref{fig:scaling} demonstrates strong scaling of {\sofa} up to 44 Summit nodes (264 GPUs) for relatively large datasets, i.e.,
  AU, DI, PR, and HM. Particularly, {\sofa} achieves speedups of 31.0$\times$, 24.9$\times$, 21.5$\times$, and 16.1$\times$, respectively, for AU, DI, PR, and HM. It is worth noting that {\sofa} can effectively use a number of GPUs that is not necessarily a power-of-two, which provides great flexibility to the application code.}
  We also notice that the HM and PR enjoy close to linear scalability from 1 to 33 nodes but flat out from 33 to 44 nodes because these two datasets do not have the adequate workload to saturate 44 nodes.

\begin{figure}[h!]
	\centering
	\includegraphics[width=.98\textwidth]{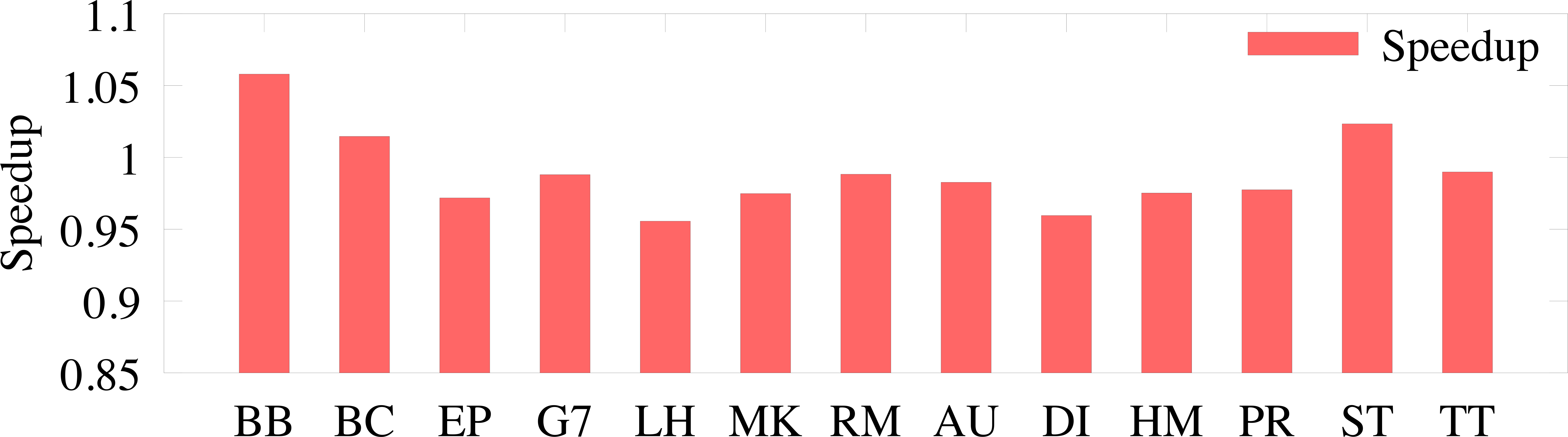} 
	\vspace{-.075in}
	\caption{{Impacts of using line 9.5 in algorithm in Figure}~\ref{fig:alg}.
	  \vspace{-.075in}
	} 
	\label{fig:max_fill_lead} 
\end{figure}

\textbf{Performance impacts of optimizing access to maxId(:) and fill(:)}. {Putting fill(:) ahead of maxId(:) introduces observable performance differences. Particularly, as shown in Figure}~\ref{fig:max_fill_lead} {in a GPU, enabling line 9.5 in Figure}~\ref{fig:alg} {leads to performance gain in BB, BC and ST datasets with the maximum gain of 5.5\% on BB dataset. This performance gain indicates that noticeable fill-ins are repeatedly detected. Likewise, among remaining datasets, LH dataset experiences a maximum performance drop of 4.6\% because the number of maxId(:) update after fill detection of a vertex is small. Given the majority of the datasets experience performance drop with the enabling of line 9.5, we disable line 9.5 for all the datasets in our evaluations.}

\begin{figure}[h]
	\centering
	\includegraphics[width=.95\textwidth]{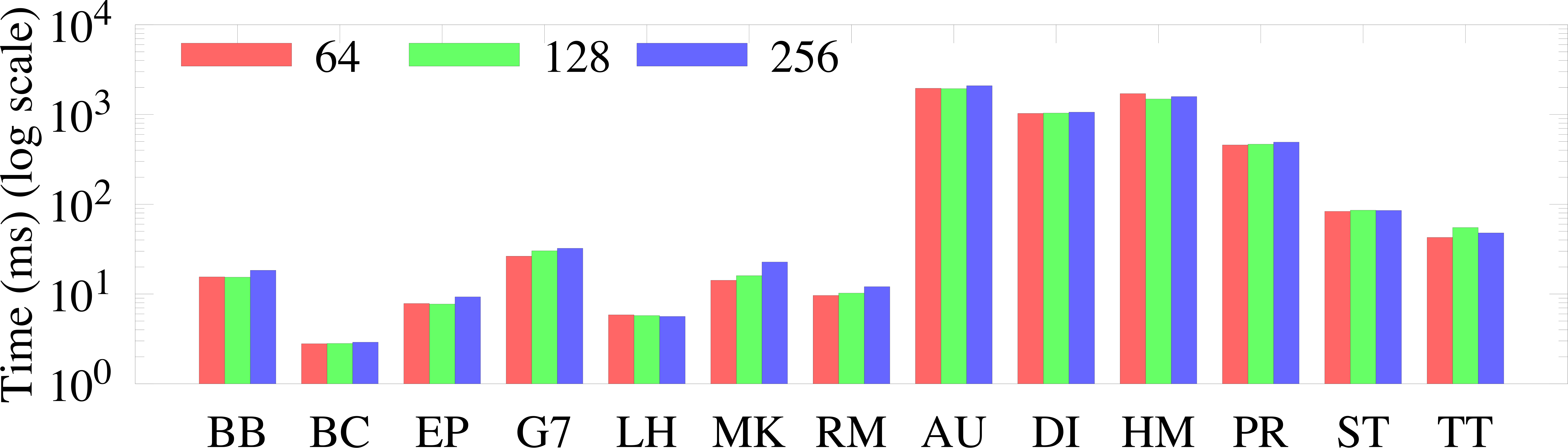} 
	\vspace{-.1in}
	\caption{{\sofa} performance variation with respect to \textit{chunkSize} = 64, 128 and 256 in six Summit compute nodes. 
	  \vspace{-.2in}
	} 
	\label{fig:diffChunkSize} 
\end{figure}

\textbf{{\sofa} performance variation with \textit{chunkSize.}}
{Figure}~\ref{fig:diffChunkSize} {presents the performance impact of \textit{chunkSize} from Equation}~\ref{eq:source} {to {\sofa}}.
{In general, one can observe that the increase of the \textit{chunkSize} leads to longer time. On average, the performance degradation is 1.04$\times$ and 1.13$\times$ when we increase \textit{chunkSize} from 64 to 128 and 256, respectively, with the maximum slowdown as 1.6$\times$ of MK for \textit{chunkSize} = 64 to 256. The reason behind this trend is that the increase of the \textit{chunkSize} typically leads to more imbalanced workload. In this work, we follow SuperLU\_DIST to set \textit{chunkSize} = 128 for {\sofa} as the default configuration even though chunkSize = 64 is slightly faster.}

\section{Related Work}\label{sec:related}

Most of the major state-of-the-art sparse LU factorization codes, such as GLU3.0~\cite{peng2020glu3},~\cite{ren2012sparse},~\cite{he2015gpu} and SuperLU\_DIST~\cite{li2003superlu_dist} adopt the  Gilbert-Peierls algorithm~\cite{gilbert1988sparse} for symbolic factorization. 
GLU3.0,~\cite{ren2012sparse} and~\cite{he2015gpu} directly use the sequential version of the Gilbert-Peierls algorithm~\cite{gilbert1988sparse}.
SuperLU\_DIST implements a parallel CPU-based symbolic factorization algorithm~\cite{grigori2007parallel} based upon Gilbert-Peierls algorithm  with symmetric pruning~\cite{eisenstat1992exploiting} and supernodal traversal~\cite{demmel1999supernodal}. To improve parallelism {for symbolic factorization}, SuperLU\_DIST resorts to nested dissection to partition $\mathbf{A}$ into independent rows {at a level of the separator tree~\cite{grigori2007parallel}}. 
{The filled structures computed at a level of the separator tree are communicated among the required processes that perform symbolic factorization at
the higher level of the separator tree.}
This leads to  inter-process communications {while the computation changes the level along the separator tree.}  To the best of our knowledge, {\sofa} is the first GPU-based parallel symbolic factorization algorithm.


It is also worth mentioning that NVIDIA cuSOLVER library~\cite{cusolver} provides solvers for sparse and dense linear systems with different approaches including LU decomposition. However, cuSOLVER does not yet support GPU version of sparse LU decomposition. 
Hence, we cannot compare  {\sofa} against the symbolic factorization of cuSolver.

Fill2 algorithm is similar to Dijkstra's Single Source Shortest Path algorithm{~\cite{dijkstra1note}} in the sense that fill2 uses the maximum vertex ID on a path to represent the ``distance'' metric in Dijkstra's algorithm. The salient difference between these two algorithms lies in that one does not need to  reduce the ``distance'' further if a fill-in is already detected in symbolic factorization. We also would like to point out that the previously proposed $\Delta$-step optimization~\cite{meyer2003delta} for Dijkstra's algorithm is not effective here: because fill2 only allows vertices that are smaller than the source to be active, $\Delta$-step will further restrict the parallelism. However, the similarity between these two algorithms suggests that our design of fine-grained symbolic factorization, external GPU optimizations, and supernode detection can potentially provide performance enhancement and space saving technique to the multi-source Dijkstra's algorithm~\cite{sao2020supernodal}. 

\section{Conclusion and Future Work}\label{sec:conclusion}
This paper introduces {\sofa} the first, to the best of our knowledge, GPU-based sparse LU symbolic factorization system. In particular, we revamp the fill2 algorithm to enable fine-grained symbolic factorization, redesign supernode detection to expose massive parallelism, balanced workload and minimal inter-node communication, and introduce a three-pronged space optimization to handle large sparse matrices. Taken together, we scale {\sofa} up to 264 GPUs with unprecedented performance. 
As the future work, we plan to integrate {\sofa} into the state-of-the-art SuperLU\_DIST package. 

\section{Acknowledgement}
This  work  was  in  part  supported  by NSF CRII Award No. 2000722, CAREER Award No. 2046102 and the Exascale Computing  Project  (17-SC-20-SC),  a  collaborative  effort  of  the  U.S. Department of Energy Office of Science and the National Nuclear Security Administration (NNSA). Any opinions, findings and conclusions or recommendations expressed in this material are those of the authors and do not necessarily reflect the views of DOE, NNSA or NSF.

\begin{spacing}{0.9}
{
\bibliographystyle{unsrt100.bst}
\bibliography{main}

\begin{thebibliography}{10}

\bibitem{davis2016survey}
Timothy~A Davis, et~al.
\newblock {A Survey of Direct Methods for Sparse Linear Systems}.
\newblock {\em Acta Numerica}, 2016.

\bibitem{grigori2007parallel}
Laura Grigori, et~al.
\newblock {Parallel Symbolic Factorization for Sparse LU with Static Pivoting}.
\newblock {\em SIAM Journal on Scientific Computing}, 2007.

\bibitem{lezar2010gpu}
E~Lezar et~al.
\newblock {GPU-based LU Decomposition for Large Method of Moments Problems}.
\newblock {\em Electronics letters}, 2010.

\bibitem{sao2014distributed}
Piyush Sao, et~al.
\newblock {A Distributed CPU-GPU Sparse Direct Solver}.
\newblock In {\em Euro-Par}. Springer, 2014.

\bibitem{liu2016synchronization}
Weifeng Liu, et~al.
\newblock {A Synchronization-free Algorithm for Parallel Sparse Triangular
  Solves}.
\newblock In {\em Euro-Par}. Springer, 2016.

\bibitem{rose1978algorithmic}
Donald~J Rose et~al.
\newblock {Algorithmic Aspects of Vertex Elimination on Directed Graphs}.
\newblock {\em SIAP}, 1978.

\bibitem{gilbert1993elimination}
John~R Gilbert et~al.
\newblock {Elimination Structures for Unsymmetric Sparse LU Factors}.
\newblock {\em SIMAX}, 1993.

\bibitem{gilbert1988sparse}
John~R Gilbert et~al.
\newblock {Sparse Partial Pivoting in Time Proportional to Arithmetic
  Operations}.
\newblock {\em SISC}, 1988.

\bibitem{dijkstra1959note}
Edsger~W Dijkstra.
\newblock {A Note on Two Problems in Connexion with Graphs}.
\newblock {\em Numerische mathematik}, 1959.

\bibitem{demmel1999supernodal}
James~W Demmel, et~al.
\newblock {A Supernodal Approach to Sparse Partial Pivoting}.
\newblock {\em SIMAX}, 1999.

\bibitem{liu2016ibfs}
Hang Liu, et~al.
\newblock {ibfs: Concurrent Breadth-first Search on GPUs}.
\newblock SIGMOD, 2016.

\bibitem{duff1999design}
Iain~S Duff et~al.
\newblock {The Design and Use of Algorithms for Permuting Large Entries to the
  Diagonal of Sparse Matrices}.
\newblock {\em SIMAX}, 1999.

\bibitem{tinney1967direct}
William~F Tinney et~al.
\newblock {Direct Solutions of Sparse Network Equations by Optimally Ordered
  Triangular Factorization}.
\newblock {\em Proceedings of the IEEE}, 1967.

\bibitem{cuthill1969reducing}
Elizabeth Cuthill et~al.
\newblock {Reducing the Bandwidth of Sparse Symmetric Matrices}.
\newblock In {\em 24th National Conference}, 1969.

\bibitem{kernighan1970efficient}
Brian~W Kernighan et~al.
\newblock {An Efficient Heuristic Procedure for Partitioning Graphs}.
\newblock {\em The Bell system technical journal}, 1970.

\bibitem{karypis1998parallel}
George Karypis et~al.
\newblock {A Parallel Algorithm for Multilevel Graph Partitioning and Sparse
  Matrix Ordering}.
\newblock {\em Elsevier}, 1998.

\bibitem{nvidia2017nvidia}
Tesla NVIDIA.
\newblock {NVIDIA Tesla V100 GPU Architecture}, 2017.

\bibitem{li2019compiler}
Lingda Li et~al.
\newblock {Compiler Assisted Hybrid Implicit and Explicit GPU Memory Management
  Under Unified Address Space}.
\newblock In {\em SC}, 2019.

\bibitem{sparse_matix}
Suite Sparse~Matrix Collection.
\newblock {Suite Sparse Matrix Collection.}
\newblock Retrieved from \url{https://sparse.tamu.edu/ }, 2020.
\newblock Accessed: 2019, November 15.

\bibitem{li2003superlu_dist}
Xiaoye~S Li et~al.
\newblock {{SuperLU\_DIST}: A Scalable Distributed-memory Sparse Direct Solver
  for Unsymmetric Linear Systems}.
\newblock {\em TOMS}, 2003.

\bibitem{parmetis}
Karypsis Lab.
\newblock {ParMETIS.}
\newblock Retrieved from \url{http://glaros.dtc.umn.edu/gkhome/metis/parmetis/
  }.
\newblock Accessed: 2020, June 02.

\bibitem{liu2015enterprise}
Hang Liu et~al.
\newblock {Enterprise: Breadth-first Graph Traversal on GPUs}.
\newblock In {\em SC}, 2015.

\bibitem{merrill2012scalable}
Duane Merrill, et~al.
\newblock {Scalable GPU Graph Traversal}.
\newblock {\em Acm Sigplan Notices}, 47(8):117--128, 2012.

\bibitem{gaihre2019xbfs}
Anil Gaihre, et~al.
\newblock {XBFS: Exploring Runtime Optimizations for Breadth-first Search on
  GPUs}.
\newblock HPDC, 2019.

\bibitem{cudaMemAdvise}
NVIDIA.
\newblock {Cuda Toolkit Documentation.}
\newblock Retrieved from \url{https://docs.nvidia.com/cuda/index.html }.
\newblock Accessed: 2020, August 11.

\bibitem{summit}
Oak Ridge~National Laboratory.
\newblock {Summit: America's newest and smartest supercomputer}.
\newblock Retrived from \url{https://www.olcf.ornl.gov/summit/}.
\newblock Accessed: 2018, August 6.

\bibitem{peng2020glu3}
Shaoyi Peng et~al.
\newblock {GLU3.0: Fast GPU-based Parallel Sparse LU Factorization for Circuit
  Simulation}.
\newblock {\em IEEE Design \& Test}, 2020.

\bibitem{wang2016gunrock}
Yangzihao Wang, et~al.
\newblock {GUNROCK: A High-performance Graph Processing Library on the GPU}.
\newblock In {\em PPoPP}, 2016.

\bibitem{ren2012sparse}
Ling Ren, et~al.
\newblock {Sparse LU Factorization for Parallel Circuit Simulation on GPU}.
\newblock In {\em DAC}, 2012.

\bibitem{he2015gpu}
Kai He, et~al.
\newblock {GPU-accelerated Parallel Sparse LU Factorization Method for Fast
  Circuit Analysis}.
\newblock {\em IEEE VLSI Systems}, 2015.

\bibitem{eisenstat1992exploiting}
Stanley~C Eisenstat et~al.
\newblock {Exploiting Structural Symmetry in Unsymmetric Sparse Symbolic
  Factorization}.
\newblock {\em SIMAX}, 1992.

\bibitem{cusolver}
{CUDA Toolkit Documentation}.
\newblock Available at \url{https://docs.nvidia.com/cuda/cusolver/}.

\bibitem{dijkstra1note}
Edsger~W Dijkstra et~al.
\newblock {A Note on Two Problems in Connexion with Graphs}.
\newblock {\em Numerische mathematik}, 1959.

\bibitem{meyer2003delta}
Ulrich Meyer et~al.
\newblock {$\Delta$-stepping: A Parallelizable Shortest Path Algorithm}.
\newblock {\em Journal of Algorithms}, 2003.

\bibitem{sao2020supernodal}
Piyush Sao, et~al.
\newblock {A Supernodal All-pairs Shortest Path Algorithm}.
\newblock PPoPP, 2020.

\end{thebibliography}
}
\end{spacing}





%
\begin{wrapfigure}{r}{0.15\textwidth}
  \begin{center}
   	\includegraphics[width=.4\textwidth]{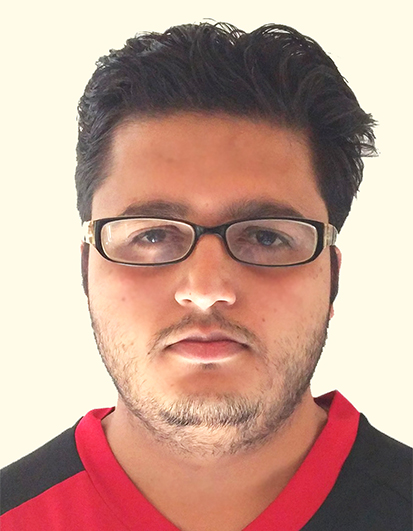} 
  \end{center}
\end{wrapfigure}
 
\vspace{-.4in}
\begin{IEEEbiographynophoto}{Anil Gaihre}
is a Ph.D. candidate at the Department of Electrical and Computer Engineering, Stevens Institute of Technology. His research interests include graph theory, sparse linear algebra, high performance computing in multi-core computer architectures and blockchain. 
His prior work experience involves working as a lead Software Engineer at E\&T Nepal Pvt. Ltd. that involved Research and Development on CFD simulations on GPUs.  
\end{IEEEbiographynophoto}

\vspace{-4mm}
\begin{wrapfigure}{r}{0.15\textwidth}
  \begin{center}
   	\includegraphics[width=.4\textwidth]{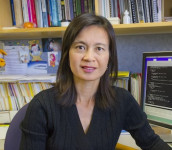} 
  \end{center}
\end{wrapfigure}
\begin{IEEEbiographynophoto}{Xiaoye Sherry Li}
is a Senior Scientist at Lawrence Berkeley National Laboratory. She has worked on diverse problems in high performance scientific computations, including parallel computing, sparse matrix computations, high precision arithmetic, and combinatorial scientific computing. She has (co)authored over 120 publications, and contributed to several book chapters. She is the lead developer of SuperLU, a widely-used sparse direct solver, and has contributed to the development of several other mathematical libraries, including ARPREC, LAPACK, PDSLin, STRUMPACK, and XBLAS. She is a SIAM Fellow and an ACM Senior Member.
\end{IEEEbiographynophoto}

\vspace{-4mm}
\begin{wrapfigure}{r}{0.15\textwidth}
  \begin{center}
   	\includegraphics[width=.4\textwidth]{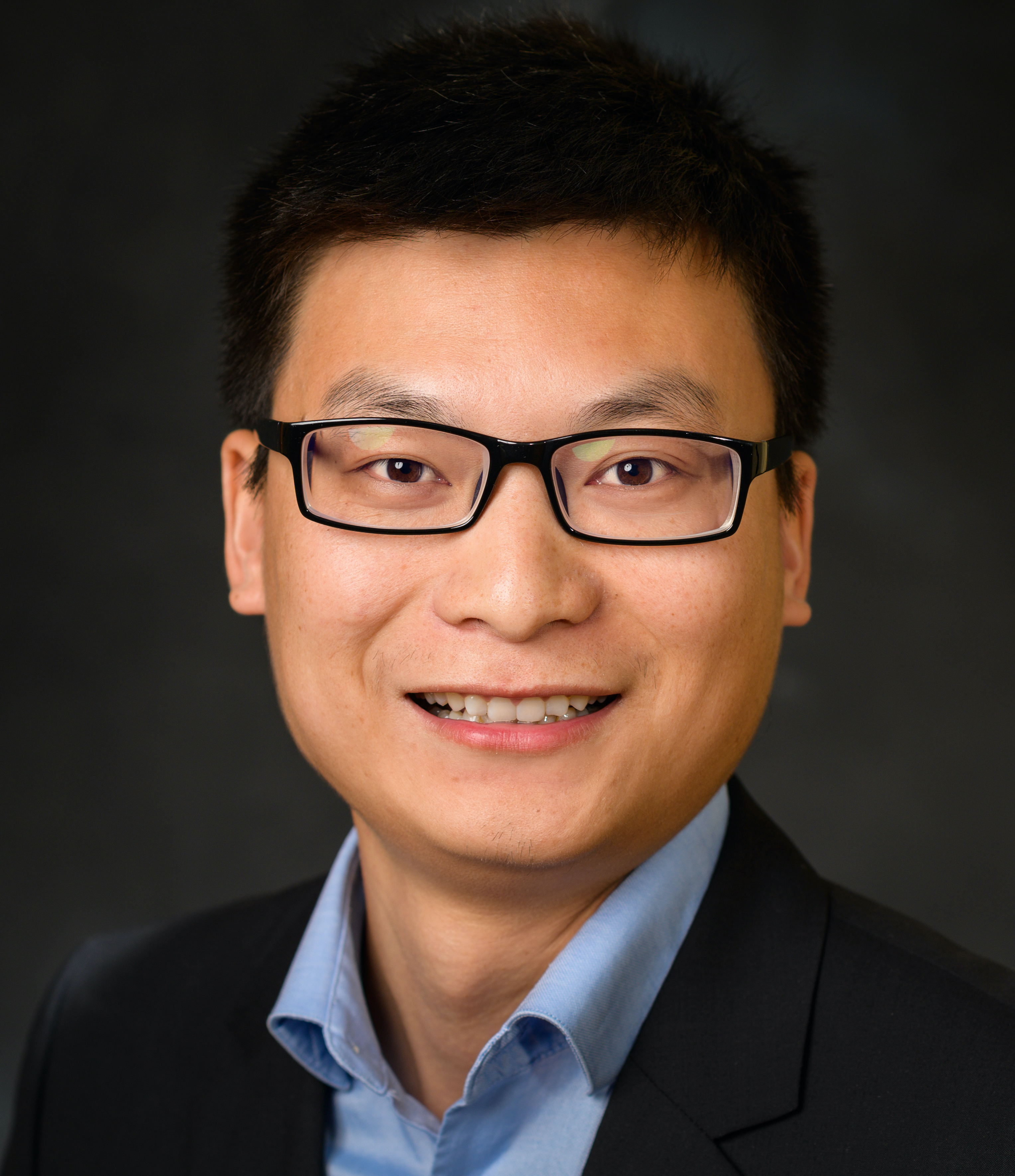} 
  \end{center}
\end{wrapfigure}
\begin{IEEEbiographynophoto}{Hang Liu}
is an Assistant Professor of Electrical and Computer Engineering at Stevens Institute of Technology. Prior to joining Stevens, he was an assistant professor at the Electrical and Computer Engineering Department of University of Massachusetts Lowell. He is on the editorial board for Journal of BigData: Theory and Practice, a program committee member for SC, HPDC, and IPDPS, and regular reviewer for TPDS and TC. He earned his Ph.D. degree from the George Washington University 2017. He is the Champion of the MIT/Amazon GraphChallenge 2018 and 2019, and one of the best papers awardee in VLDB '20. 
\end{IEEEbiographynophoto}

\newpage
\setcounter{page}{1}

Supplement File for {\sofa}: Scalable Sparse LU Symbolic Factorization on GPU.

\section{Examples for LU decomposition Phases}\label{app:other_phases}

\begin{figure}[hbt!]
	\centering
	\includegraphics[width=.9\textwidth]{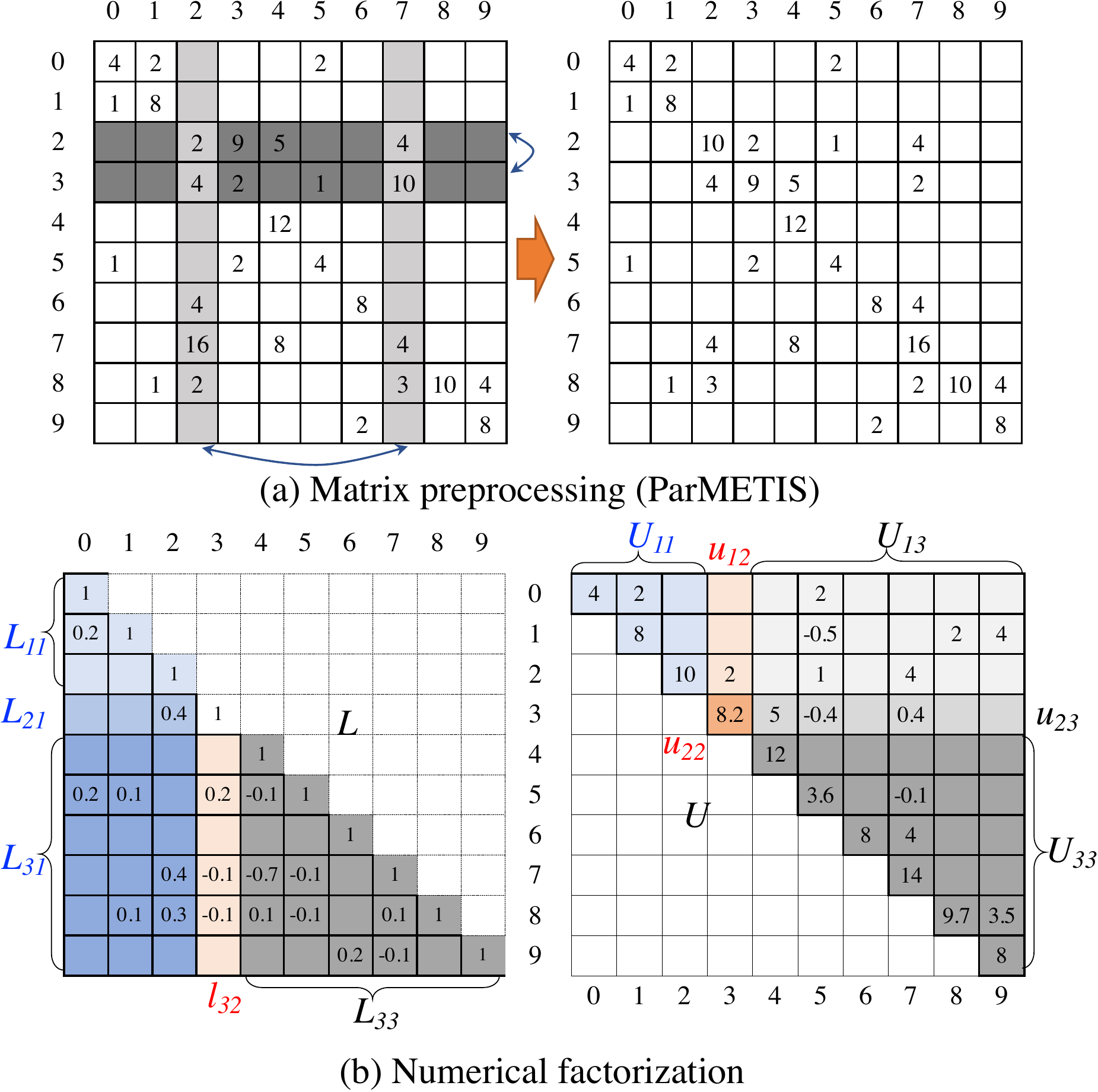} 
	\caption{Other phases in LU decomposition.} 
	\label{fig:diff_phases_detail} 
\end{figure}

\textbf{Matrix preprocessing.} 
Figure~\ref{fig:diff_phases_detail}(a) demonstrates the matrix preprocessing on an example matrix. Particularly, we swap columns 2 and 7 so that column 2 will have 4 instead of 5 nonzeros. Further, we perform a swap between rows 2 and 3 for better numerical stability, i.e., larger numerical values are moved to diagonal. 

\textbf{Numerical factorization.}
Figure~\ref{fig:diff_phases_detail}(b) shows a left-looking approach of numerical factorization to determine the k-th column of the filled matrix by using the computed results from columns 0 to k-1. Using column 3 of ($\mathbf{L}$+$\mathbf{U}$) in Figure~\ref{fig:diff_phases_detail}(b) as an example, we will use columns 0, 1 and 2 to solve this column. Representing $\mathbf{A}$, $\mathbf{L}$ and $\mathbf{U}$ in block-matrix form, we can write LU factorization of A as follows.

\begin{equation} \label{eq:leftlooking}
\setlength\arraycolsep{3pt}
    \begin{bmatrix}
    \footnotesize
\textcolor{mygreen}{L_{11}} &  & \\
\textcolor{mygreen}{l_{21}} & 1 & \\
\textcolor{mygreen}{L_{31}} & \textcolor{red}{l_{32}} & L_{33}\\
\end{bmatrix} 
    \begin{bmatrix}
\textcolor{mygreen}{U_{11}} &  \textcolor{red}{u_{12}}& U_{13}\\
 & \textcolor{red}{u_{22}} & u_{23}\\
 &  & U_{33}\\
\end{bmatrix} = \begin{bmatrix}
\textcolor{mygreen}{A_{11}} &  \textcolor{mygreen}{a_{12}}& \textcolor{mygreen}{A_{13}}\\
\textcolor{mygreen}{a_{21}} & \textcolor{mygreen}{a_{22}} & \textcolor{mygreen}{a_{23}}\\
\textcolor{mygreen}{A_{31}} & \textcolor{mygreen}{a_{32}} & \textcolor{mygreen}{A_{33}}\\
\end{bmatrix},
\end{equation} 

where \textcolor{mygreen}{$L_{11} = L(0:2, 0:2)$}, \textcolor{mygreen}{$l_{21} = L (3:3, 0:2)$}, \textcolor{mygreen}{$L_{31} = L(4:9,0:2)$}, \textcolor{red}{$l_{32} = L(4:9, 3:3)$}, $L_{33} = L(4:9, 4:9)$, \textcolor{mygreen}{$U_{11} = U(0:2, 0:2)$}, \textcolor{red}{$u_{12} = U (0:2, 3:3)$}, $U_{13} = U(0:2, 3:9)$, \textcolor{red}{$u_{22} = U (3:3, 3:3)$}, $u_{23} = U (3:3, 4:9)$, and $U_{33} = U(4:9, 4:9)$. 
The texts in \textcolor{mygreen}{green}, \textcolor{red}{red} and black colors represent the \textcolor{mygreen}{known}, \textcolor{red}{currently under solving}, and unknown blocks, respectively. Through block matrix multiplication of Equation~(\ref{eq:leftlooking}) towards $a_{12}$, $a_{22}$ and $a_{32}$, we further obtain:
\begin{equation}
\setlength\arraycolsep{3pt}
\label{eq:leftlooking_solve}
  \begin{aligned}
  \textcolor{mygreen}{L_{11}}{\textcolor{red}{u_{12}}} = \textcolor{mygreen}{a_{12}},  \\
 \textcolor{mygreen}{l_{21}}\textcolor{red}{u_{12}} + {\textcolor{red}{u_{22}}} = \textcolor{mygreen}{a_{22}}, \\
 \textcolor{mygreen}{L_{31}}\textcolor{red}{u_{12}} + {\textcolor{red}{l_{32}}}\textcolor{mygreen}{u_{22}} = \textcolor{mygreen}{a_{32}},
  \end{aligned}
\end{equation}
where \textcolor{red}{$u_{12}$}, \textcolor{red}{$u_{22}$} and \textcolor{red}{$l_{32}$} are computed in order in Equation~(\ref{eq:leftlooking_solve}) so that the \textcolor{red}{$u_{22}$} and \textcolor{red}{$l_{32}$} can be calculated once \textcolor{red}{$u_{12}$} is resolved in the first equation of Equation~(\ref{eq:leftlooking_solve}).

\end{document}